\documentclass[a4paper]{article}
\usepackage[margin=25mm]{geometry}
\usepackage[english]{babel}
\usepackage[utf8]{inputenc}
\usepackage{amsmath}
\usepackage{amsfonts}
\usepackage{graphicx}
\usepackage{latexsym}
\usepackage{graphics}
\usepackage{caption}
\usepackage{cite}

\providecommand{\keywords}[1]
{
  \small	
  \textbf{\textit{Keywords---}} #1
}

\title{\textbf{Socioeconomic biases in urban mixing patterns of US metropolitan areas}}
\author{Rafiazka Millanida Hilman$^{1}$, Gerardo Iñiguez$^{1,2,3}$, Márton Karsai$^{1}$  \\
        \small $^{1}$Department of Network and Data Science, Central European University, 1100 Vienna, Austria  \\
        \small $^{2}$Department of Computer Science, Aalto University School of Science, 00076 Aalto, Finland \\
        \small $^{3}$Centro de Ciencias de la Complejidad, Universidad Nacional Autónoma de México, 04510 Ciudad de México, Mexico \\
}
\date{} 

\begin{document}
\maketitle

\begin{abstract}
Urban areas serve as melting pots of people with diverse socioeconomic backgrounds, who may not only be segregated but have characteristic mobility patterns in the city. While mobility is driven by individual needs and preferences, the specific choice of venues to visit is usually constrained by the socioeconomic status of people. The complex interplay between people and places they visit, given their personal attributes and homophily leaning, is a key mechanism behind the emergence of socioeconomic stratification patterns ultimately leading to urban segregation at large. Here we investigate mixing patterns of mobility in the twenty largest cities of the United States by coupling individual check-in data from the social location platform Foursquare with census information from the American Community Survey. We find strong signs of stratification indicating that people mostly visit places in their own socioeconomic class, occasionally visiting locations from higher classes. The intensity of this `upwards bias' increases with socioeconomic status and correlates with standard measures of racial residential segregation. Our results indicate an even stronger socioeconomic segregation in individual mobility than one would expect from system-level distributions, shedding further light on uneven mobility mixing patterns in cities.  
\end{abstract} \hspace{10pt}


\section{Introduction}
\label{into}
\setlength{\parindent}{2em}
\setlength{\parskip}{1em}
Patterns of socioeconomic inequality can be found everywhere in a modern city. Large variations in earned income leading to uneven access to services, healthcare and education~\cite{rumberger2005does,acevedo2003residential}, as well as spatial and housing segregation~\cite{taeuber2008residential,iceland2002racial}, are just two of the most drastic examples of socioeconomic disparity. Less studied is the segregation related to mobility mixing, where people from different socioeconomic classes encounter each other less often than what is potentially allowed by the city fabric~\cite{dong2020segregated,moro2019atlas,netto2015segregated}.

Big data presents a unique opportunity to analyse the role of human mobility in segregation, from the level of individuals to the scale of societies. Digital data tracing human movements in cities ranges from mobile call detail records (CDRs) \cite{gonzalez2008understanding,song2010modelling} and GPS trajectories \cite{tang2015uncovering,gallotti2016stochastic,alessandretti2018evidence}, to location-sharing services (LSS) and check-in sequences on social media platforms \cite{hawelka2014geo,wu2014intra,jurdak2015understanding}. The analysis of these data sources, providing anonymised individual trajectories with unprecedented spatiotemporal resolution, has proven essential for our growing understanding of the underlying mechanisms of human mobility \cite{brockmann2006scaling,baronchelli2013levy,wang2014correlations,alessandretti2020scales}, and the associated ability to predict future trajectories \cite{beiro2016predicting,comito2018human}. It also offers the possibility to engage in a more comprehensive and nuanced exploration of urban socioeconomic segregation, by combining high-dimensional mobility data with information on the socioeconomic traits of individuals \cite{luo2016explore,leo2016socioeconomic,dong2020segregated}.

Earlier studies on human mobility present evidence of characteristic spatial scales \cite{brockmann2006scaling,gonzalez2008understanding,song2010modelling,alessandretti2020scales}, as well as a correlation between human spatial behaviour and socioeconomic dynamics \cite{marston2000social,paasi2004place,boterman2016cocooning}. Rather than being homogeneously mixed, human mobility (as represented by daily individual trajectories throughout urban spaces) is strongly influenced by socioeconomic preferences. People sharing socioeconomic backgrounds are more likely to visit similar places within their class range and interact amongst themselves \cite{bora2014mobility,yip2016exploring,wang2018urban,morales2019segregation}, thus generating stratified mobility and social network patterns. In the presence of homophily mixing~\cite{mcpherson2001birds}, spatial exploration is dictated by one's socioeconomic class, reducing the number of visits to locations with different economic status, and thus inducing highly predictable trajectories. However, when people aspire to diversify their experiences by, e.g., visiting lavish areas of the city, where they have never been able go before, the potential for an upwards bias in visiting patterns appears. Meanwhile, other has studied the effects of segregation of mixing in urban places using location data to explore exploration/exploitation behavioural patterns and their correlations with socioeconomic status~\cite{moro2021mobility}.

Homophily mixing is not the only mechanism influencing mobility patterns. The variability of socioeconomic traits such as ethnic group, education level, occupation sector, etc. also constrains the possibility of movement in urban spaces via residential segregation~\cite{taeuber2008residential,iceland2002racial}, where people with similar backgrounds live next to each other and form fragmented areas in the city. Given the potentially complex interplay between human mobility and socioeconomic stratification, it is worth asking whether the presence of biased mobility across tracts of some socioeconomic trait is associated with lower residential segregation. This is particularly relevant given the number of studies reporting mobility as a key pillar in diminishing segregated spaces among people from diverse groups in society \cite{schonfelder2003activity,wong2011measuring,farber2013social,farber2015measuring}. People show heterogeneity in many aspects, including their mobility characteristics and socioeconomic capacities, which shape their patterns of movement across urban space. In an attempt to better understand such mobility patterns, we focus here on socioeconomic and mobility heterogeneities at the individual and class levels. Given the heavy-tail nature of human mobility (in which a few outliers have remarkably longer trajectories and higher socioeconomic diversity of visited places), heterogeneities may vary across levels. Therefore, it is in our interest to quantify the variability of individual mobility and to compare it with the typical behaviour of others in a given socioeconomic class.    

In this study, we emphasise the need to query the extent to which behavioural segregation (bias in mobility) is related to residential segregation. We take a step forward in the current analysis of segregation in mobility by asking the following question: How do socioeconomic attributes and geographic constraints affect the spatiotemporal process of individuals moving in urban spaces? To answer this question, we analyse individual check-in trajectories in the twenty largest cities of the United States, coupled with detailed socioeconomic maps indicating the economic status of people and places they visit. After a short data description, in the following we will introduce stratification matrix measures and individual- and class-level mobility bias scores to quantify patterns of mobility segregation, visiting biases, and their variation across cities with wide-ranging socioeconomic and ethnics segregation profiles.

\section{Data description}
\label{data}
\setlength{\parindent}{2em}
\setlength{\parskip}{1em}

In order to simultaneously capture the mobility patterns and socioeconomic status of people, we concentrate on two independent sources (mobility and socioeconomic data, described below) and combine them using spatial information.

\emph{Mobility data:} To construct individual mobility trajectories, we analyse a large, open Foursquare dataset~\cite{yang2016participatory}, which records how people move from one place to another. Data comes as a sequence of user check-ins to places, or points of interest (POIs), thus providing information on mobility trajectories of individuals and visiting frequencies of places. This dataset is not collected directly through the Foursquare open API, but from Foursquare check-ins via Twitter. The crawling method corresponds to 18 months of observations between April 2012 and September 2013 for users with Foursquare-tagged tweets. Using this mobility data, constituted by roughly 37,860 people with nearly 1,857,100 check-ins, we concentrate only on active users (who checked-in from at least two different places during the observation period). Focusing on the 20 largest metropolitan areas in the US, we also infer the home locations of 26,502 users following a conventional pipeline of conditions~\cite{mcneill2017estimating} [for a detailed description of the method and a statistical summary see Supplementary Material (SM) Section A].

\begin{figure*}[!htb]
 \centering
    \includegraphics[width=1.\linewidth, keepaspectratio]{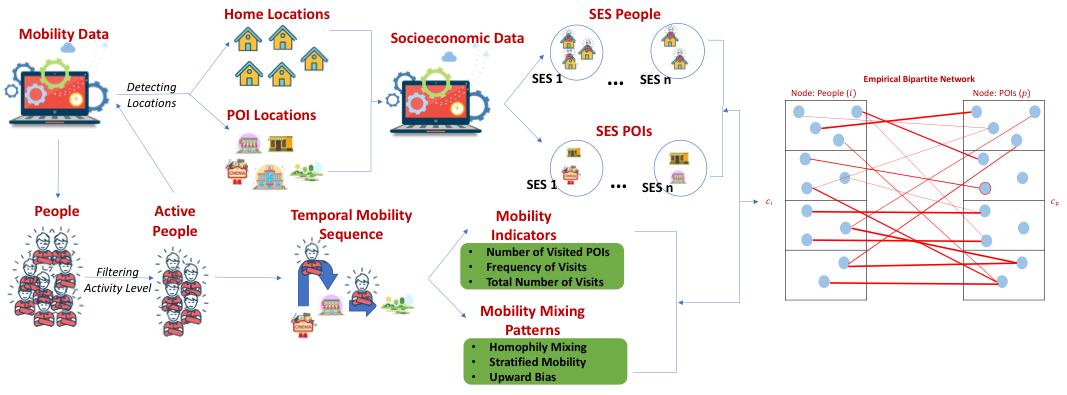}
	\caption{\textbf{Mobility and socioeconomic data combination pipeline.} {\bf (left)} Overview of data sources, data processing pipelines and data combination steps to obtain data for the analysis of socioeconomic segregation in spatiotemporal urban mobility. {\bf (right)} As a result we obtain a bipartite network, with nodes classified into two sets comprising individuals \(u \) and POIs \(p \). Each node in both types is labelled by a socioeconomic indicator ($c_{U}$ and $c_{P}$) assigned via our location-based method on the census tract level. Weighted edges between individuals and POIs indicate the frequency of visits of a given user to a given place.} 
\label{fig:1}       
\end{figure*}

\emph{Socioeconomic data:} To estimate the socioeconomic status of people and places, we rely on the 2012 American Community Survey (ACS)~\cite{census2012} (recorded in the year matching the closest to the Foursquare observation period). After identifying the corresponding ACS census tract where a user's home location lies, we associate the socioeconomic indicators of this location to the individual. In order to estimate the economic status of a place, we follow a similar strategy and assign local socioeconomic status indicators to POIs based on their locations. Although the socioeconomic status of venues could arguably be better estimated from their pricing, this information is at present not available to us. Thus, we assume that the socioeconomic status (SES) of people living at a location is well correlated with the pricing of venues in the same neighbourhood and offered services around (for a summary of our data construction pipeline see Figure~\ref{fig:1}).

In order to obtain a proper representation of socioeconomic status in the context of segregation, we consider 78 features from the ACS data. Although such a large number of dimensions in principle provides a rich way of quantifying the socioeconomic status of locations (and people living there), it turns out these variables have high redundancy. We perform a principal component analysis to identify the most relevant ones and find that income features (11 variables) have the largest loading, accounting for most of the socioeconomic variance between places. After implementing three different techniques (mutual information rank~\cite{kraskov2004estimating}, decision tree~\cite{friedman2001elements}, and Gini coefficient~\cite{raileanu2004theoretical}), per capita income consistently stands out as the best indicator of individual SES: It accounts for the largest variance and it correlates strongly with other income variables such as earning/wage, wealth, and supplementary source of income (for more details on this analysis see SM Section A). By using the average per capita income as the socioeconomic indicator of active users living in a given tract, we sort them in an ascending order. To group them into distinct socioeconomic classes, we then segment this sorted list into 10 equally populated groups with people of the lowest income in class 1 and highest income in class 10. By means of this procedure we assign a socioeconomic class $c_U$ to each user. In identical fashion, each venue is assigned a value $c_P$.

\section{Results}
\label{sec:res}
\setlength{\parindent}{2em}
\setlength{\parskip}{1em}

Our main scientific goal is to study socioeconomic segregation and biases in population mixing in cities by observing correlation patterns between the SES of people and places they visit. Using the collected data, this objective can be addressed by building a network of individuals visiting places. We define a stratified bipartite network $G=(U,P,E)$, where individual $u$ is a node in set $U$, and place $p$ belongs to set $P$ (with $U\cap P=\emptyset$). People and places are connected by edges $e_{u,p}\in E$ with weights $w_{u,p}$ coding the number of times person $u$ visited place $p$ (see Figure~\ref{fig:1}). Further, we stratify $U$ into a set of socioeconomic classes indexed by values from $C_U$ thus assigning a class membership $c_u=i\in C_U$ to each individual.

In the same way we define $c_p=j\in C_P$ classes for places. This network representation captures all information about the socioeconomically stratified visiting patterns of people to venues, coding their possible encounters and giving an aggregated description of the potential mixing patterns of people of different socioeconomic classes.

\subsection{Matrix measures}

\begin{figure*}[ht!]
    \centering
        \includegraphics[width=0.75\linewidth]{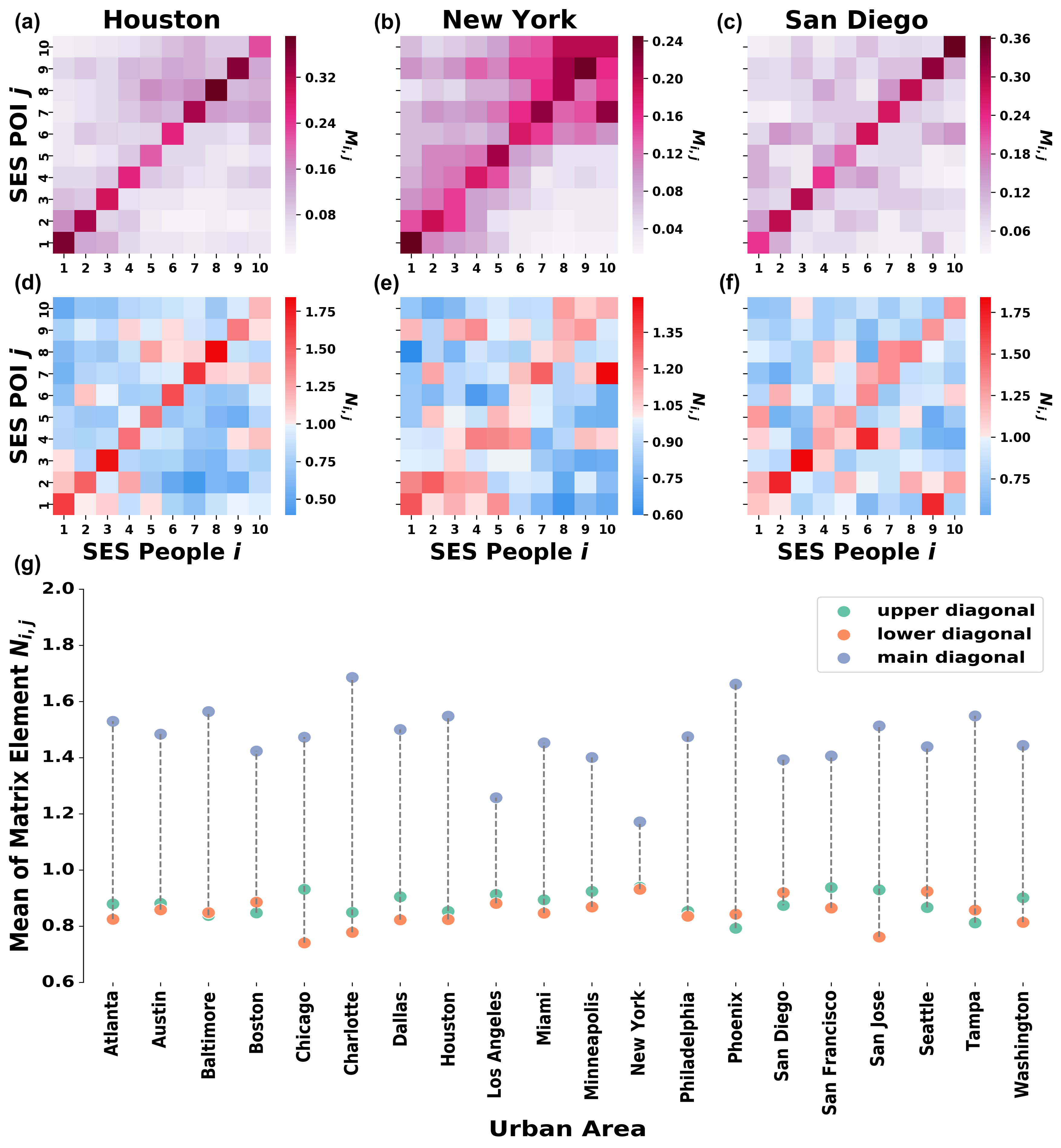}
        \caption{\textbf{Socioeconomic stratification matrices.}  
        (a) The empirical stratification matrices $M_{i,j}$, showing the probabilities that individuals from a given class visit to places of different classes. The darker colour shades of bins represent larger visiting probability. Matrices of Houston (Fig.~\ref{fig:2}a), New York (Fig.~\ref{fig:2}b) and San Diego (Fig.~\ref{fig:2}c) all show strong stratification patterns, indicating that people tend to visit most likely places with similar status.
        (b) The normalised stratification matrices $N_{i,j}$, defined as the fraction of the empirical and randomised stratification matrices. After normalisation, such stratification pattern becomes less evident for New York (Fig.~\ref{fig:2}e) and San Diego (Fig.~\ref{fig:2}f) but quite persistent in Houston (Fig.~\ref{fig:2}d). Similar matrices computed for other urban areas are available in SM Section B and C. (c) Mean of matrix element $N_{i,j}$, computed separately for upper diagonal, lower diagonal, and main diagonal. Among 20 urban areas, 13 of them including Houston and New York have higher mean values for upper diagonal elements (Fig.~\ref{fig:2}g), indicating dominant upward visiting biases. In contrast, dominant downward visiting biases are found in San Diego. 
        }
        \label{fig:2}       
\end{figure*}

Based on the bipartite network representation we can measure the frequency at which people of a given class visit places in different classes. To summarise these visiting patterns we use stratification matrices~\cite{leo2016socioeconomic}. An \emph{empirical stratification matrix} gives the probability that a person $u\in U$ from a given socioeconomic class $c_u=i\in C_U$ visits a place $p\in P$ belonging to a class $c_p=j\in C_P$. More formally:
\begin{equation}
M_{i,j}= \frac{\sum_{U,c_u=i}\sum_{P,c_p=j} w_{u,p}}{\sum_{j\in C_P}\sum_{U,c_u=i}\sum_{P,c_p=j} w_{u,p}},
\end{equation}

where the numerator counts the number of times people from class $i$ visit places of class $j$, and the denominator normalises this frequency matrix column-wise to obtain a visiting probability distribution for each individual class $i\in C_U$. Such matrices are shown in Figure~\ref{fig:2} for selected cities (Houston, New York and San Diego). The dominant diagonal elements for Houston and San Diego indicate strongly stratified visiting patterns in these cities. People prefer to visit places of their own or similar socioeconomic class, rather than places from remote classes. Interestingly, for New York this pattern is less evident suggesting weaker socioeconomic preferences in visiting venues.

To decide if these patterns appear as the consequence of population statistics or other confounding effects, we compare the matrix $M_{i,j}$ to a reference matrix, which measures similar stratification patterns in a system where visiting patterns appear uniformly at random with certain constraints. This \emph{randomised stratification matrix} is defined through a random rewiring process of the bipartite network, while constraining the total number and frequency of visits of each individual (i.e. their activity and link weights), the class of individuals and places, but fully randomising links between individuals and visited places otherwise. The randomisation is performed by selecting randomly for each link of an individual $u$ a place to visit from the set of places ever visited by their respected socioeconomic classes $c_u$, while keeping the link weight intact. This \emph{in-class randomisation} allows us to compare an individual's behaviour to similar others, meanwhile distinguishing between socioeconomic classes, which potentially are characterised by very different visiting patterns.

After generating randomised bipartite networks via 100 independent realisations, we compute a similar column-wise normalised stratification matrix $R_{i,j}$, representing the probability of people from class \(c_i \) randomly visiting to places of class \(c_p \). To finally obtain whether the empirical mobility patterns appear more frequently than by chance, we compare the empirical and the preference-based null model matrices. We obtain a \emph{normalised stratification matrix} $N_{i,j}$ by taking the element-by-element fraction of the empirical and random contact matrices as: 
\begin{equation}
N_{i,j}=\frac{M_{i,j}}{R_{i,j}}.
\end{equation}
Each element in the matrix $N_{i,j}$, which are \(N_{i,j}\) $>$ 1 (red bins in Fig.~\ref{fig:2}d-f) indicates that the visits made by individuals from class $i\in C_U$ to place of class $j\in C_P$ appeared with higher probability in the empirical observations than it was expected from the random null model. 

Otherwise, the blue blocks for \(N_{i,j}\) $<$ 1 show that the corresponding visits appeared with a smaller probability than expected by chance. In cases red bins dominate the diagonal of the normalised matrix $N_{i,j}$, it indicates patterns of socioeconomic stratification, where people prefer to visit places of similar socioeconomic status as their own, rather than places, which are richer or poorer than them. This is the case of Houston and San Diego (see Fig.~\ref{fig:2}d and f respectively) and many other cities listed in SM Section B. However, this character is less evident for New York (see Fig.~\ref{fig:2}e), where despite known strong residential segregation, the city fabric mitigates a more homogeneous mixing of people.

These normalised stratification matrices reveal further characters of possible biases of people in choosing places to visit, out of their own class. If in a city people exhibit upward visiting biases, thus they tend to choose more expensive places to visit when they step out of their own class, the upper diagonal matrix elements of $N_{i,j}$ would appear dominantly red. While, if the opposite is true, the lower diagonal elements would reflect similar but downward visiting biases. To simply quantify these patterns, we compute the average values $N_{i,j}$ elements of normalised stratification matrix of cities above, at, and under their diagonals. From Fig.~\ref{fig:2}g it is clear that, in all cities, diagonal elements dominantly concentrate visiting probabilities. However, in terms of off-diagonal averages, in most of the cities (like in Houston in Fig.~\ref{fig:2}d) the upper diagonal average takes a larger value as compared to the lower diagonal average, indicating present upward visiting biases in these metropolitan areas. Meanwhile, in some cases the contrary is true (like in San Diego in Fig.~\ref{fig:2}f) or in some cities these averages are very similar thus indicating no dominant upward or downward visiting biases, as in case of New York (see Fig.~\ref{fig:2}e and Fig.~\ref{fig:2}g).

\subsection{Individual bias}

The matrix measures presented in Fig.~\ref{fig:2} reflect the coexistent socioeconomic configurations derived from visit trajectories. Firstly, the empirical stratification matrices $M_{i,j}$ bring an initial indication of homophily mixing as seen in the dominant frequency visit within own class. Secondly, these results reveal the underlying inclination in visiting places situated in higher SES as depicted by the larger proportion of upper diagonal elements in the normalised stratification matrices $N_{i,j}$ in most of the cities. Taking these two configurations into account, it can be inferred that while individual mobility is dictated by the membership of socioeconomic class most of the time, the embedded motivation to visit upper class places is still present. 

We take a technical step ahead in order to adequately quantify this visiting bias that indicates deviations in mixing from the respected $c_{u}$ socioeconomic class of an individual. We compute a single \emph{empirical individual bias score} for each individual $u\in U$ as
\begin{equation}
B_u= \langle c_p\rangle_u - c_{u},
\label{eq:Bi}
\end{equation}
where $\langle c_p\rangle_u=\frac{\sum_{p\in P} w_{u,p}\times c_{p}}{n_p^u}$ is the average socioeconomic status of places an individual $u$ visited, defined as the fraction of the $\sum_{p\in P} w_{u,p}\times c_{p}$ sum of socioeconomic status of places in the trajectory of individual $u$ and the $n_p^u=\sum_{p\in P} w_{u,p}$ number of times individual $u$ visited any places. An individual has upward visiting bias if her individual score $B_u$ is positive, meaning that she tends to visit places located in more affluent areas than where she lives. Secondly, an individual with negative score value has downward visiting bias since places she usually visits are situated in lower socioeconomic class than her own. Otherwise, an individual does not have any indication of bias ($B_u=0$) if she visits places within her own socioeconomic rank. A reference model for this measure can follow a similar logic as the in-class randomisation for the realisations of network reference models explained before. Given the individual trajectory resulted from the random visit generating process, we calculated a \emph{randomised individual bias score} using the same formula as in Eq.~\ref{eq:Bi}. Note that in this measure boundary effects may appear, as people from the poorest class cannot exhibit downward bias, and similarly, the highest class cannot be upward biased. Individual bias scores can be fairly compared to null models, which retain these boundary effects. In-class randomisation fulfils this requirement, providing an average randomised bias score $\langle B^{rand} \rangle_u$ for each individual separately. Note, that the randomised individual bias scores take non-trivial values, different from zero, due to the individual variance of visiting frequencies of individuals to different places. These are represented by the weights $w_{u,p}$ in the bipartite network, which are preserved during the randomisation process.

The comparison of the empirical and in-class normalised individual bias scores can be best quantified by an individual bias z-score as 
\begin{equation}
z_u^{B_u}=\frac{B_u-\langle B^{rand} \rangle_u}{\sigma_{u}^{B^{rand}}},
\end{equation}
where $\langle B^{rand} \rangle_u$ is the mean and $\sigma_{u}^{B^{rand}}$ is the standard deviation of the randomised individual bias scores across 100 independent realisations of the null model. The value of $Z_u^{B_u}$ reflects how much the individual bias deviates from the expected bias for an individual who chooses places to visit with the same frequency as before but selects them from a given set of places dictated by others within the same socioeconomic class.

\begin{figure*}[!htb]
    \centering
    \includegraphics[width=0.95\linewidth]{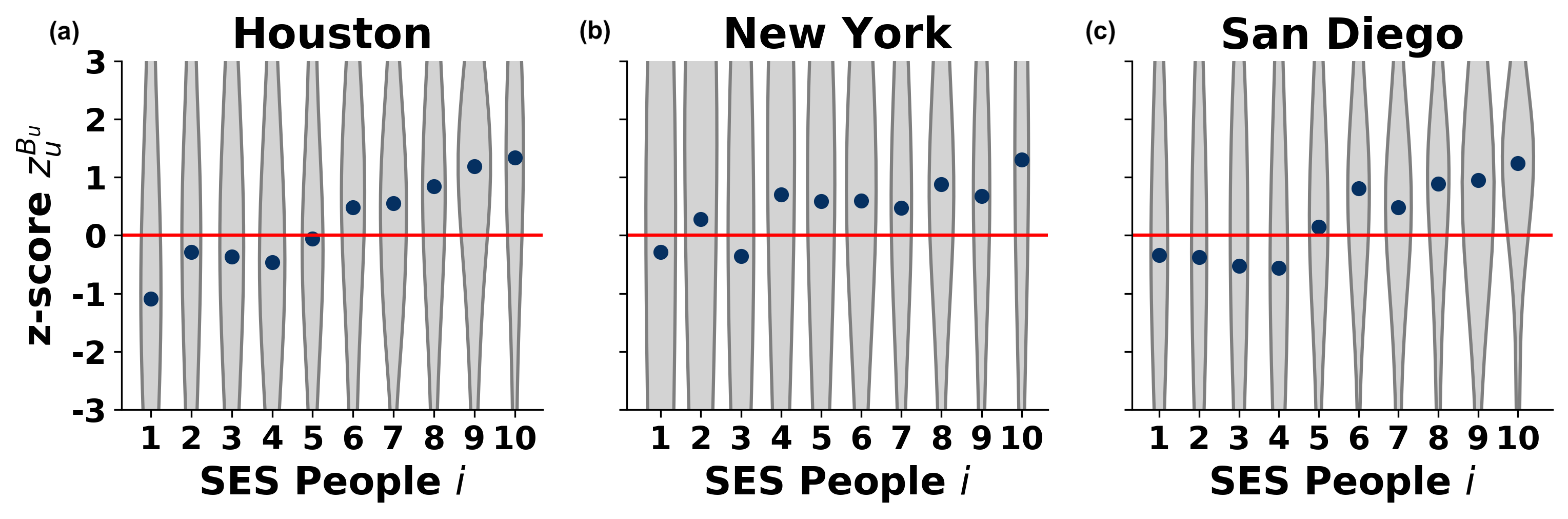}
    \caption{\textbf{Individual Bias z-score $z_u^{B_u}$}. Class level distributions and their median values are shown for each socioeconomic class in Houston (Fig.~\ref{fig:3}a), New York (Fig.~\ref{fig:3}b) and San Diego (Fig.~\ref{fig:3}c). The overall increasing trend of medians (blue dots) indicates that people from lower classes are less biased than expected, while the contrary is true for others from higher classes. Solid red line indicates the fully unbiased case. For results on other cities see Section C and Fig.6 in the SM.} 
\label{fig:3}
\end{figure*}

The class distributions of individual z-scores together with their median values are shown in Fig.~\ref{fig:3}, where the unbiased level is assigned as a flat red line. These distributions appear broad for each class, indicating that actually people from any class exhibit upward or downward biases in terms of their visiting patterns to other socioeconomic classes. Interestingly, the median z-scores indicate an increasing trend in all the three depicted cities. The people from lower classes appear with slightly negative bias z-score, meaning they have a slightly weaker bias to visit places of different socioeconomic classes than expected from their random visiting patterns. In contrary, the middle and upper classes are evidently biased stronger than expected. This increasing trend of the median of the individual bias z-score with socioeconomic classes surprisingly characterises all the investigated cities as shown in Section C and in Fig.6 in the SM. Although this is a robust pattern, certain variations appear between cities, which can be explained by the spatial distribution of venues with respect to the socioeconomic fabric of the city population.

\subsection{Class-level bias}

The individual bias score $B_u$ compares the average class of visited places of an individual to its own socioeconomic rank inferred from its home location. Meanwhile, its z-score $z_u^{B_u}$ indicates if this individual bias is weaker or stronger than expected from random behaviour. However, this measure is using the class label of the individual as a reference of comparison, and it says less about whether an individual visits higher or lower class places as compared to the random expected behaviour characterising other individuals in its own class. To directly measure this effect we introduce a class level z-score measure
\begin{equation}
z_{u}^{c_u}=\frac{\langle c_p \rangle_u - \langle c_p \rangle_{c_u}^{rand}}{\sigma_{c_u}^{rand}},
\end{equation}
where $\langle c_p \rangle_u$ is the average socioeconomic status of places individual $u$ visited, and $\langle c_p \rangle_{c_u}^{rand}$ and $\sigma_{c_u}^{rand}$ are the average and standard deviation (respectively) of class of places that others from class $c_u$ would visit if behave randomly. This reference measure, just like before, is generated by in-class shuffling to obtain null models over $100$ realizations. The value of $z_{u}^{c_u}$ reflects directly how much the individual behaviour deviates from the expected level, when the individual could choose randomly places to visit from a given set dictated by others from the same socioeconomic class.

\begin{figure*}[!htb]
    \centering
        \centering
        \includegraphics[width=0.95\linewidth]{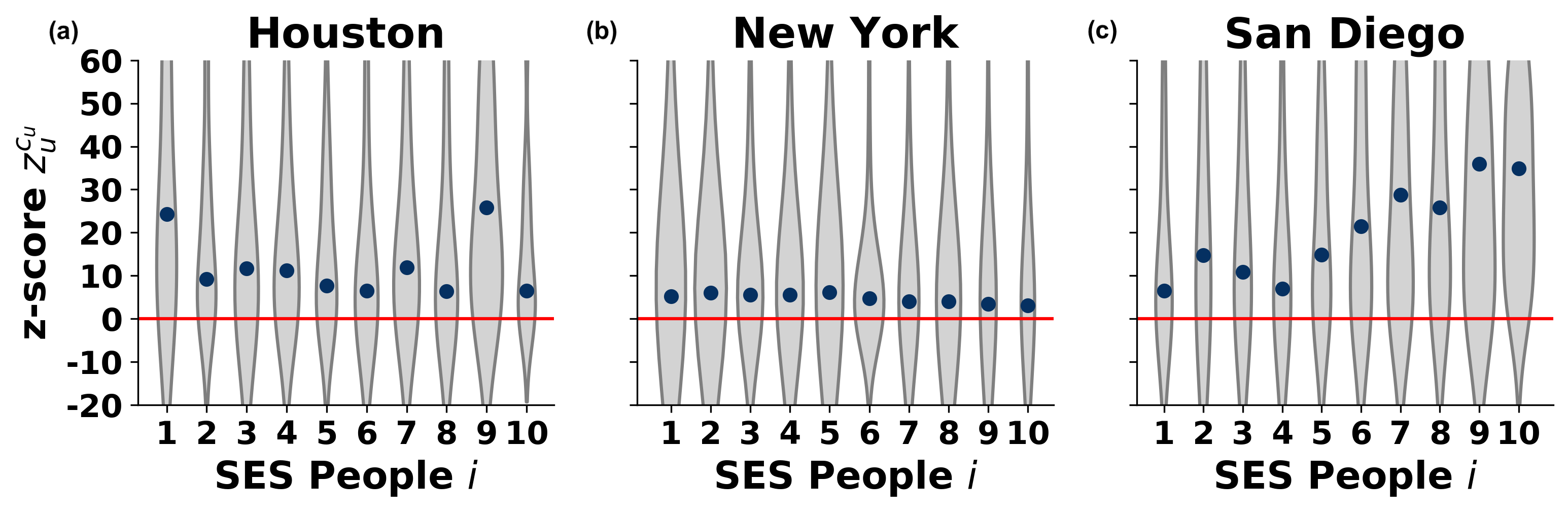}
        \caption{\textbf{Class-level Bias z-score $z_{u}^{c_u}$}. Distribution of class-level biased z-scores as the function of socioeconomic classes. Distributions are shown for each socioeconomic class with their median values as blue points for Houston (Fig.~\ref{fig:4}a), New York (Fig.~\ref{fig:4}b), and San Diego (Fig.~\ref{fig:4}c). Z-score values corresponding to unbiased cases are shown with red solid lines. Positive z-score values signal an upward visiting bias characterizing each city. For results on other cities see Section C and Fig.7 in the SM.}
\label{fig:4}       
\end{figure*}

Results in Fig.~\ref{fig:4} show a different behaviour as compared to the individual bias scores. In case of New York (see Fig.~\ref{fig:4}b), the distributions of the class level bias z-scores indicate that, although the variation is large in each socioeconomic classes, the medians of these distributions are all slightly positive and independent of the socioeconomic class. This signals a weak upward bias in people's visiting patterns in New York as compared to the class behaviour that appears for each class. In other cities, we find several other bias patterns during our analysis (see SM Section C and Fig.7). In case of San Diego (in Fig.~\ref{fig:4}c) class-level biases are all positive and evidently increasing with the socioeconomic classes. This suggests that richer people in San Diego may visit even more affluent places, than one would expect from their random class behaviour. Somewhat the opposite trend can be observed for Houston (Fig.~\ref{fig:4}a), where although the class-level bias z-score is always positive and indicates upward bias for each class, it seems to follow an overall decreasing trend.

\begin{figure}[!htb]
\centering
\includegraphics[width=0.8\textwidth]{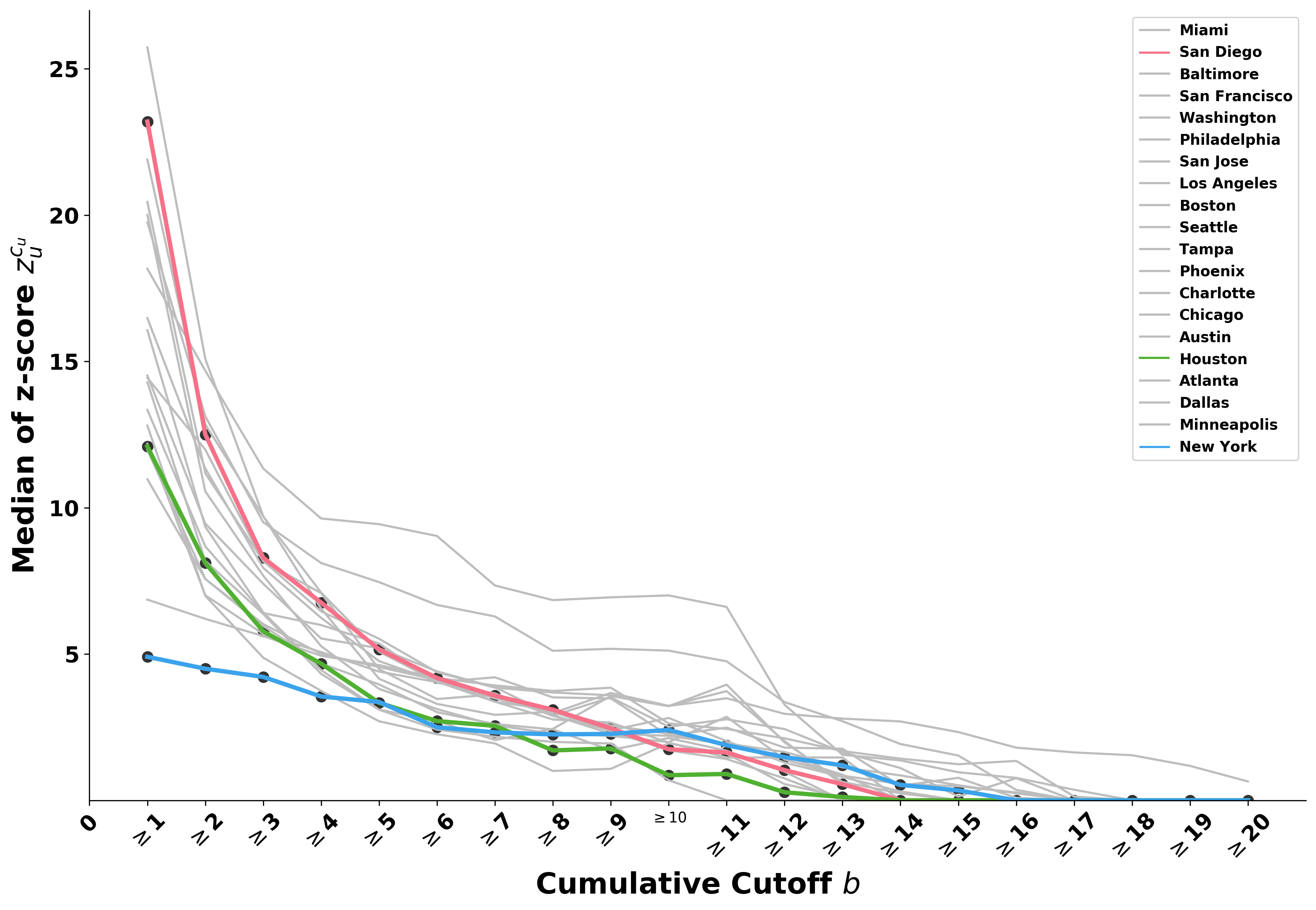} 
\caption{\textbf{Sensitivity of class-level bias z-score $z_{u}^{c_u}$}. Lower bound cutoff is set as $b\geq1$ to which we only take into account venues visited at least once. For each set of venues in individual trajectory cumulatively visited $b$ times or higher, we measure class-level bias z-score $z_{u}^{c_u}$ and take the median values. Upper bound cutoff $b\geq20$ is added to accommodate venues visited even more frequently. As $b$ incrementally becomes larger, the medians are largely dropped closer to 0. It indicates that venues visited more frequently tends to be more homogeneous in term of mixing and closer to own socioeconomic status.}
\label{fig:5}
\end{figure} 

Visiting patterns measured by the class-level bias scores suggest that an upward socioeconomic bias characterise each cities we study. However, although these measures incorporate the visiting frequency distribution of individuals, they do not show evidently that upward biases typically appear due to repeated visits to places with higher class scores, or due to several occasional visits to places out of ones socioeconomic classes. To answer this question, we recompute the median bias scores, excluding places which were visited less number of times by an individual than a given threshold. Results are depicted in Fig.~\ref{fig:5} for each city. As expected, the median class-level bias score appears as a decreasing function of the frequency threshold in each city. This suggests that people visit more frequently places, which are closer in terms of socioeconomic status to their own class, while visit more affluent places occasionally only, that in turn causes upward bias patterns characterising their class. Beyond this general decreasing character, this function indicates large variance between different cities. For example, in case of San Diego (red line in Fig.~\ref{fig:5}), this curve starts from a high z-score value when all visits are consider but decreases rapidly as repeated visits are taken into account. In case of New York this function starts from a relatively small z-score values and decrease linearly for larger threshold values. This suggests a different visiting behaviour where people typically visit places more than one time, but closer to their own socioeconomic class.

\subsection{Mobility mixing and segregated residences}

While there is an expected relation between the mobility mixing patterns and residential segregation in a city, the combined investigation of these phenomena has not received much attention so far. Their relation is important however for several reasons. For example, due to the multitude correlated socioeconomic factors it is likely that e.g. ethnicity, which strongly correlates with income status in US metropolitan areas, correlates also with residential segregation, as it has been shown in several studies~\cite{iceland2002racial, abramovitz2021persistence}. On the other hand, the daily mobility of people and their visiting patterns to different places are constrained also by these socioeconomic factors, thus they are likely to resemble similar segregation patterns. To investigate these correlations, we focus on different ethnic groups and the likelihood of their mixing in cities, which exhibits different level of mobility segregation patterns.

To quantify the level of segregation in mobility mixing, we analyse the earlier introduced normalised stratification matrix $N_{i,j}$ for each city. As we have discussed, signatures of segregation can be associated to strong diagonal elements in these matrices, indicating that people of a given SES are the most likely to visit places associated with the same or similar SES, as compared to random visiting patterns. To quantify the strength of diagonal concentration of visiting probabilities, we measure the diagonality index of the normalised stratification matrices~\cite{bokanyi2021universal}, which is similar to the assortativity coefficient used by others~\cite{dong2020segregated,newman2003mixing}. It is defined as the Pearson correlation coefficient of matrix entries as
\begin{equation}
\ r_N = \frac{\sum_{i,j} i j N_{i,j} - \sum_{i,j}i N_{i,j}\sum_{i,j} j N_{i,j}}{\sqrt{\sum_{i,j} i^2 N_{i,j} - \left(\sum_{i,j} i N_{i,j}\right)^2}{\sqrt{\sum_{i,j} j^2 N_{i,j} - \left(\sum_{i,j} j N_{i,j}\right)^2}}}.
\end{equation}
Here $i\in c_u$ indicates the socioeconomic class of individuals and $j\in c_p$ is the same for places. The diagonality index takes values between $-1$ and $1$. In case it is $1$, it indicates perfect assortative mixing corresponding to a fully stratified matrix with non-zero elements in its diagonals and zero anywhere else. Cities with large $r_n$ values are characterised by visiting patterns of people who are strictly bounded to places associated to their own socioeconomic class. On the contrary, if $r_n$ takes smaller than zero values (in extremity $r_n=-1$), it indicates dis-assortative connections between people and places of different socioeconomic status. This corresponds to mobility mixing patterns where people prefer to visit places of different SES rather than places from their own class. In case $r_n=0$, the normalised stratification matrix is flat indicating no choice preferences of people to visit places with particular SES.

The mixing patterns in a city may not be only determined by the socioeconomic status of people but also by residential segregation. Residential segregation is strongly correlated with the ethnicity of people ~\cite{massey1988dimensions, logan2011persistence, abramovitz2021persistence}, which in turn, according to Wang and others~\cite{wang2018urban}, is an even stronger predictor of mobility mixing than socioeconomic status when it turns to black, Hispanic, and white poor and non-poor populations. This study found that the minority groups - despite their socioeconomic status - have lower exposure to richer or white neighborhoods, comparing to poor white groups. The fact that they travel across similar distance and frequency to many places, does not change the persistent pattern of their isolation and segregation. Therefore, racial segregation emerges from a higher-order level, not limited to their residential neighborhood but expanded to their mobility and potential contact. 

To address the effects of residential segregation on mobility mixing, we took a similar path than others~\cite{krivo2013social,wang2018urban} and considered the ethnic group distribution in a city as a proxy. Residential segregation is indicated by housing clustering tendency of individuals from the same ethnic group. This can be formally quantified by the so-called distance decay isolation~\cite{morgan1983distance}, which measures the probability that a racial group minority interacts with members of their own group by considering the distance from the racial group minority's housing area. This is measured as:
\begin{equation}
Dp_{xx*} =\sum_{i=1}^{n}\left(\frac{x_i}{X}\sum_{j=1}^{n}\frac{k_{ij}x_j}{t_j}\right),
\end{equation}
where $x_i$ and $x_j$ are the population sizes of a minority group in census tracts $i$ and $j$ (respectively),  $X=\sum_i x_i$ is the total population of the minority group, and $t_j$ is the total population of census tract $j$. The distance decay dimension is reflected by $k_{ij} = \frac{t_j^{-d_{ij}} }{\sum_{j=1}^{n}t_{j}^{-d_{ij}}}$, where $d_{ij}$ is the distance between the centroids of census tracts $i$ and $j$. Hence, higher index suggests higher probability of interaction with people from the same group, inferring isolation from the rest of population. In our case, we use a probabilistic individual profiling to identify the most likely socioeconomic profile of an individual based on the ethnic group with the highest proportion at the respected census tract where one lives. For instance, if an individual $u$ lives at census tract $i$ where the racial composition there is $60\%$ white, $15\%$ Hispanic, $10\%$ black, and $5\%$ Asian, this individual is considered as white. We consider different thresholds at first and we find out that considering a neighbourhood the ethnicity if such people consist of at least the 30\% of the given tract is the optimum cut-off because it is the highest threshold with the lowest unidentified census tract ethnicity profiles.

\begin{figure}[!h]
\centering
\includegraphics[width=.8\textwidth]{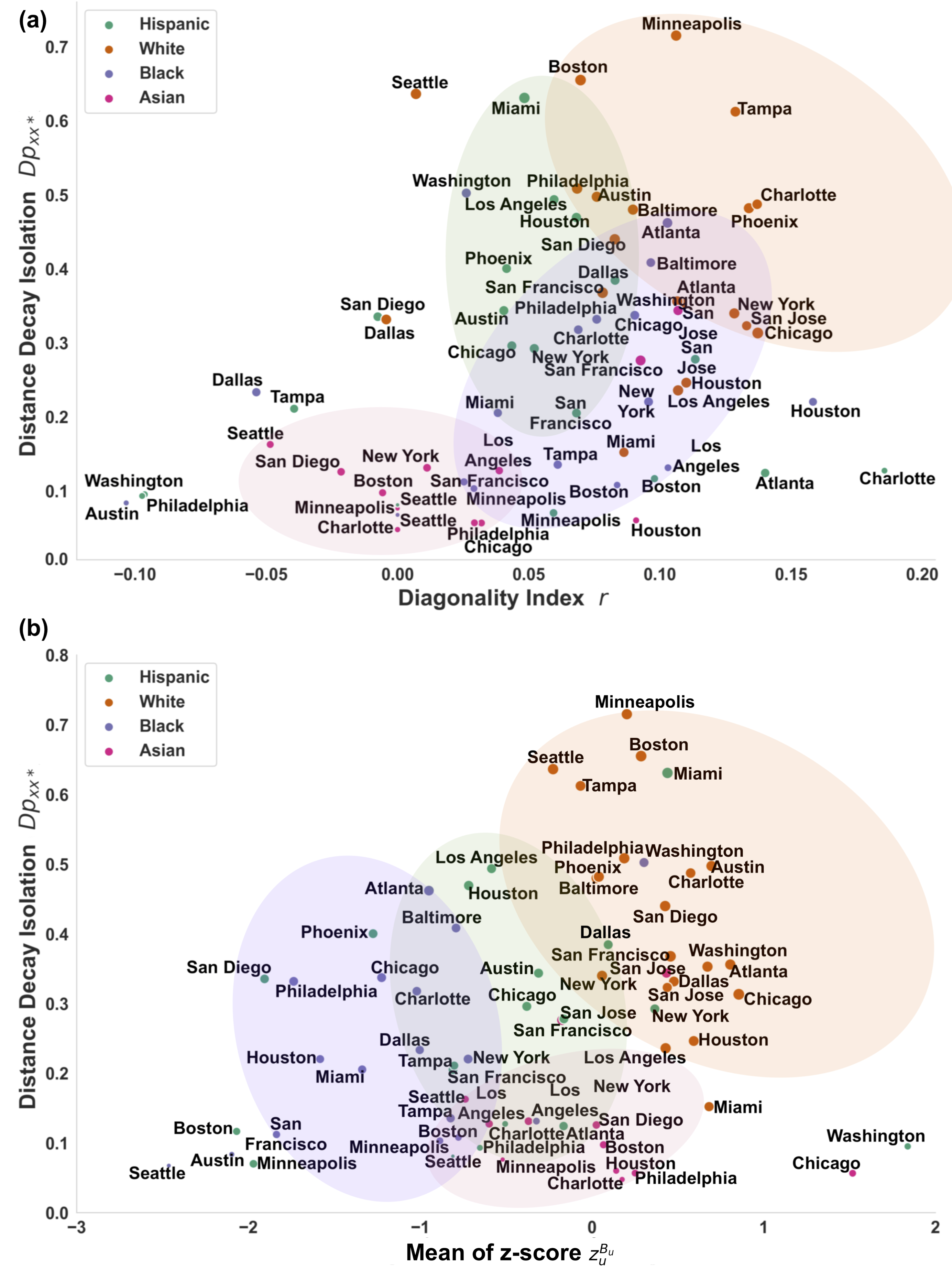} 
\caption{Segregation and bias measure correlations with isolation scores for different ethnic groups. Panel (a) depicts the correlation between the diagonality index $r_N$ and  distance decay isolation $Dp_{xx*}$ while panel (b) show a similar correlation of the average individual bias z-score $z_u^{B_u}$. In each plot colours of symbols and blobs indicate ethnic groups of Hispanic (green), White (green), Black (purple) and Asian (red) people. The sizes of symbols are scaled with the size of these ethnic population identified in the Foursquare dataset in each city. Blobs with respected colour illustrate the cluster formation based on racial groupings. The shape is arbitrary, only to demonstrate the visibility of clusters.}
\label{fig:6}
\end{figure}

Recalling the above mentioned diagonality index and individual bias, we take the average of z-score of each of these bias measures at the level of ethnic groups in every urban area and correlate them with their distance decay isolation value computed for the same ethnic group in the same city. By considering four ethnic groups (White, Hispanic, Black and Asian) we receive four data points for each cities as shown in Fig.~\ref{fig:6}. To see how much the analysed Foursquare population is representative for each ethnic group in a city see SM Section A. Although the total number of analysed individuals are not proportional to the total population of each city, the in-city fraction of different ethnic groups are similar to the census distributions.

There is a striking correlation emerging between the diagonality index (quantifying assortativity mobility mixing of each ethnic groups) and the distance decay isolation (measuring the isolation of different ethnic groups) with $R=0.35$ ($p=0.0$). Notably, almost all diagonality index measures appear with positive values suggesting assortative mixing for most ethnic groups, with a few exceptions. Further, the overall correlation suggests the intuitive picture that the stronger mobility mixing stratification patterns characterises a city (i.e. larger its corresponding diagonality index), the stronger isolation patterns emerge between its ethnic groups. In turn, it indicates that residential segregation (and thus physical proximity) play an important role in determining visiting and mixing patterns of people in a city. More interestingly, the de-coupled ethnic groups for each city show an emerging clustering, which assigns the importance of racial differences in mobility segregation. From Fig.~\ref{fig:6}a it appears that people belonging to the white ethnic group (shown as orange points in Fig.~\ref{fig:6}a) appear to be the most isolated from the rest of the population (with the largest values of $Dp_{xx*}$), while they appear with the strongest assortative mobility segregation patterns too (with the highest diagonality indices) consistently in several cities (to lead the eye we coloured this group as an orange blob in Fig.~\ref{fig:6}a). The contrary is true for the members of the Asian ethnic groups (indicated by red points and blob in Fig.~\ref{fig:6}a). In most cities they appear as the least isolated and the most dis-assortative (least stratified) ethnic group, thus mixing well with the rest of the population. In between these two groups, people from the Hispanic ethnic group (green points and blob) seem to be more isolated than people from the black ethnic group (purple points and blob) although they show comparable strength of segregation in mobility mixing, all weaker than white people.

Needless to say that the grouping patterns shown in Fig.~\ref{fig:6}a indicate overall trends only, while several exception exists for each ethnic group. For example, the Hispanic ethnic group of Charlotte appears with the strongest assortative pattern, although this group is not strongly isolated from the rest of the population. Or the black community of Austin appears with the lowest diagonality index, suggesting a strong dis-assortative mixing of these people with the rest of the population, while they also appear as one of the least isolated among any other communities.

A similar positive correlation appears in Fig.~\ref{fig:6}b with R=0.25 ($p=0.04$) between the average individual bias z-score and distance decay isolation values over all the investigated ethnic groups and cities. Moreover, ethnic groups show certain clustering trends, which suggest ethnic trends in terms of visiting bias patterns. Interestingly, white ethnic groups (shown by orange points and blob), who we have already found the most isolated, show the strongest upward bias to visit more affluent places then their own socioeconomic class. As high SES classes are populated mostly by white people, this pattern derives from our earlier observations in Fig.~\ref{fig:3}, where we find upward bias to increase with the SES of people. Their high isolation score can be explained by their upward bias towards higher socioeconomic places, which are most likely to be visited by other white people.  Meanwhile, it may also indicate that our data have an over-represented white population, as we find upward biases in all of the cities. Strikingly, other racial groups indicate negative visiting biases and lover level of isolation. This effect is the strongest for people from black racial groups all over the country, but also characterises Hispanic and Asian communities although they show more unbiased patterns, with average individual z-score values closer to $0$.

Exceptions are again interesting. The Hispanic community of Washington appears as the most upward biased ethnic group, while the black ethnic group of Seattle sits on the other end of the spectrum and being the most downward biased minority among the analysed cities. Both of these communities appear with low level of isolation. Consequently, similar to the conclusion of Wang et al.~\cite{wang2018urban}, we observe that beyond socioeconomic status, ethnicity (strongly correlated with residential segregation) is another very important factor determining mixing patterns of people.

\section{Discussion and conclusions}
\label{dis}
\setlength{\parindent}{2em}
\setlength{\parskip}{1em}

Mobility patterns are strongly determined not only by the fabric of a city but also by the socioeconomic structure of the population living there. This leads to biased mixing and segregation in mobility, which can be observed as stratification patterns in choices to visit places. We addressed this complex phenomenon via a mobility analysis of people living in the 20 largest cities in the US, and aimed to quantify segregation patterns in mobility capturing their visit patterns to places of interest. We systematically found upward-biased mobility in all cities, with some variance across metropolitan areas. In one extreme, people living in New York do not exhibit dramatic stratification in their visit patterns but visit places in all kinds of locations, rich or poor, independently of their own socioeconomic status. Meanwhile, in Houston and San Diego people are more stratified and visit places of their own socioeconomic class, and show an upward bias towards richer places to visit. We found that this upward bias, which characterises most cities analysed, is usually induced by single visits of individuals to affluent places, while most visits correspond to their own socioeconomic class. We also revealed distinct patterns of individual mobility in terms of stratified correlations between the bias magnitude and residential segregation based on spatial distribution of racial groups in urban areas. Visual representations of ethnic clusters indicate overall trends of behaviour characterising most cities, where segregated mobility is bounded together with residential segregation and broadly contributes to the portrayal of inequality.

It should be taken into account that data for a given socioeconomic class in the population might not be comparable from one city to another due to sampling in the data collection process (see SM Section A). Multiple sources of data containing digital traces of human movements with higher resolution, such as mobile phone call records and GPS trajectories, may improve the robustness of findings presented in this paper. Methodological improvements to infer individual attributes (like racial group membership) provide a direction of future research. Moreover, algorithms for probabilistic individual profiling could be developed by using machine learning techniques such as Random Forests and Support Vector Machines in the presence of ground truth information from alternative data sources. 

One potential confounding factor of the emergent stratification patterns reported here is distance, as people visit places closer to their home more frequently, thus inducing similar correlation motifs. To check the robustness of our methods in investigating segregation in mobility and biased visiting patterns and the magnitude of such distance effect, we recomputed our results on out-of-class data after excluding own census tract visits for each individual's trajectory. Even with this constraint, SES plays a considerable role in shaping mobility (comparative observations for all cities can be found in SM Section B and D, along with SM Section F Fig.12 in the case of Houston, New York, and San Diego, in contrast to Fig.\ref{fig:2} above). On the ground of visiting biases, there are some variations among cities regarding individual bias. The earlier notion of upward visiting bias is also very much present in the case of out-of-class measurements since z-score values are all positive above the red median unbiased line (complete plots are available in SM Section C and E, while a deeper exploration for Houston, New York, and San Diego is available in SM Section F Fig.13 and Fig.14). Therefore, there are no conflicting results from our methodology even after controlling for this confounding factor. Enforcing out-of-class treatment is reasonable in this context because our study aims to analyse and quantify mixing patterns and not yet look for causal links or underlying reasons of their emergence.

Segregation is not an exclusive phenomenon to the quasi-static configuration of housing settlement, but also exists in more dynamic settings such as mobility. Questions about the conceptual relations between segregated mobility and segregated residence stand still in the literature, yet relatively untapped, while scientific investigations should follow this line of inquiry. We take a step forward through empirical data-driven analysis and yield an interaction effect between both types of segregation. Individual attributes (such as racial groups) partly explain the emergence of distinct clusters, beyond income levels. Our findings also highlight the notion that inequality is multidimensional in nature. A comprehensive policy design to address this issue should entail the wider possibility of individual movement across the urban landscape to accommodate larger socioeconomic heterophily and further interaction between socioeconomic classes.




\clearpage
\begin{center}
\font\myfont=cmr12 at 20pt
\title{\textbf{{\myfont Supplementary material}}}
\end{center}

\maketitle
\setcounter{section}{0}
\setcounter{figure}{0}
\renewcommand\thesection{\Alph{section}}
\renewcommand\thefigure{\arabic{figure}}
\section{Summary statistics and method description}
\label{SM A}

\subsection{Summary statistics}
\label{SM A1}
This study focuses on the presence of mixing pattern in the largest 20 urban areas in the US. We consider urban area as the level of analysis because this densely developed territory fits very well with our objective in capturing the interaction between peoples and places. While people are represented by Foursquare users, whom also Twitter users in this case due to the crawling technique used in data collection, places takes various built in functional point of interests (POI) in accordance with Foursquare Venue Category Hierarchy. After combining the selected geographic and cartographic information from the U.S. Census Bureau (American Community Survey/ACS 2012) and Foursquare check-ins data (April 2012 - September 2013), we infer a set of active users and urban land uses (e.g.: residential, commercial, and other nonresidential) along with their socioeconomic status (SES).  


\begin{table}[ht]
\centering
\resizebox{\textwidth}{!}{\begin{tabular}{lrrrrrrr}
  \hline
Urban Area & Number of population & Average income (US\$)  &  Number of users & Average income of  users (US\$) & Number of POIs & Number of check-ins  \\ 
  \hline
 Atlanta  &  1,515,880 & 32,154 & 1,597& 	 40,880& 	 12,729& 	 117,403\\
 \hline
 Austin	& 964,309 &	 32,508 & 1,101 	& 48,145 	& 8,114 &	 67,096 \\
 \hline
 Baltimore&	 1,558,743 &	 30,275 	& 1,006& 	 33,762& 	 9,697& 	 67,528\\
 \hline
Boston	& 2,090,520 	& 39,810 &1,653& 	 49,789& 	 13,548& 	 103,395\\
 \hline
Charlotte	& 864,236 &	 31,520 & 614 	 &41,298 	 &5,579 	 &44,738\\
 \hline
Chicago	& 2,540,254 &	 28,279 & 2,318 &	 55,997 &	 18,059& 	 166,133\\
 \hline
Dallas	& 1,368,607 &	 26,425 &	 884& 	 43,966& 	 6,160& 	 49,683\\
 \hline
Houston &	 2,027,223 &	 27,149 & 	 1,063& 	 45,351& 	 9,967 &	 67,999 \\
 \hline
Los Angeles	& 2,066,612 &	 27,162 &1,284 	 &41,929 	 &10,101 	& 99,401\\
 \hline
 Miami &	 2,146,572 	& 23,599  &1,542 	 &34,445 	 &10,509 	 &74,525 \\
 \hline
 Mineapolis	& 2,074,556 &	 34,483 &1,109 	 &35,230 	 &12,784 	 &100,677\\ 
 \hline
 New York	& 7,669,696 &	 32,225 &1,353 	& 50,457& 	 31,060& 	 191,564\\
 \hline
 Philadelphia	& 2,430,063 &	 28,960 & 1,369 	& 37,958 	& 12,445 &	 100,518\\
 \hline
 Phoenix	& 1,806,374 	& 23,744 & 969& 	 28,537& 	 8,746& 	 67,922\\
 \hline
 San Diego	& 1,732,145 &	 29,692 & 1,330 	 &41,386 	& 11,783& 	 84,476\\
 \hline
 San Francisco	& 2,059,293 &	 39,878 & 2,348 	& 51,715 &	 16,353 	& 122,846\\
 \hline
 San Jose	& 1,446,254 &	 38,089 & 762 	& 42,280 	& 6,918 	& 41,701 \\
 \hline
 Seattle &	 1,316,257 &	 41,967 & 1,148 &	 43,595 &	 10,260 &	 67,523\\
 \hline
 Tampa	& 1,120,985 	& 27,904 &	 765 	& 30,888 &	 7,506 	& 55,255\\
 \hline
 Washington	& 600,566 &	 45,237 & 2,287 &	 58,226 &	 17,662 	& 166,627\\ 
   \hline
\end{tabular}}
\caption{\textbf{Summary statistics. It summarises size of dataset used in this study, including population information (eg: number and income) extracted from census as well as Foursquare users and POIs in the 20 Urban Areas}. Number of population is the summation of real population in the census tract that appears in the Foursquare dataset. Number of users is counted from pool of users with identified home locations after running home location algorithm explained in SM \ref{SM A2}. Number of check-ins is the total frequency of users tagging themselves at every listed POI in the urban area where they live.} 
\label{tab:A1}
\end{table} 

\begin{figure*}[!hb]
 \centering
    \includegraphics[width=0.5\linewidth]{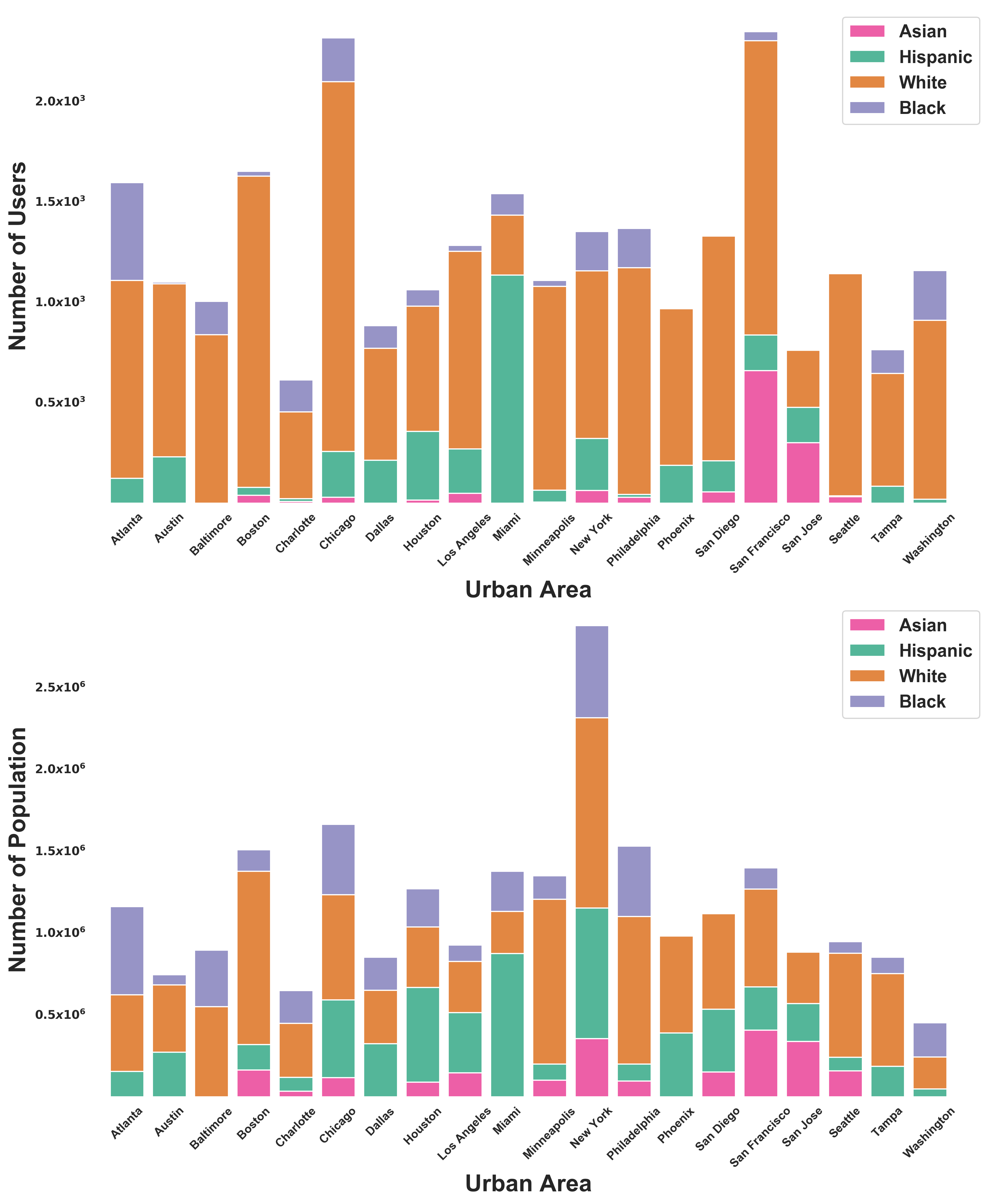}
	\caption{\textbf{Data representativeness by ethnic group}. It indicates how representative the number of users in the dataset with assigned ethnicity as compared to the census. The ratio of ethnic composition doesn't deviate much from the actual mixture at population level.} 
\label{fig:A1}       
\end{figure*}

\subsection{Home location inference}
\label{SM A2}
For each individual mobility trajectory, we identify which locations categorised as home and venues, including visits to places located in their own SES tract. Home inference algorithm is constructed based on daily temporal windows. It consists of 8 time slots starting from midnight with 4-hour intervals. The main criterion of home location is consistent check-ins from 9PM to 6AM (time window 1, 2, and 8) at locations labelled by Foursquare as 'Assisted Living', 'Home (private)', 'Housing Development', 'Residential Building (Apartment / Condo)', and 'Trailer Park'. The rank of home location candidacy is sorted based on frequency of check-ins.  We iterate the process for other venues to accommodate users that might live in the downtown, next to multi-purposes buildings attached to various urban functionalities (e.g.: a flat on the top of deli or store). This algorithm is designed with ability to properly identify the logical consequence of urban built environment where the human activities frequently take place at a location serving numerous settings as previously mentioned.

\subsection{Method description}
\label{SM A3}
\setlength{\parindent}{0pt}
In Section 2, we briefly mention the application of Principal Component Analysis (PCA) to identify the most relevant socioeconomic features. This feature extraction technique is used to narrow down the range of variables that might portray inequality and segregation at best. It reveals that the largest loading belongs to income features that in total count for 11 variables, comprising mean household income, mean earnings, median household income, median earnings for male full-time, per capita income, median earnings for female full-time, median earnings for workers, mean retirement income, mean social security income, mean supplemental security income, and mean cash public assistance income.

\begin{figure*}[!htb]
 \centering
    \includegraphics[width=0.8\linewidth]{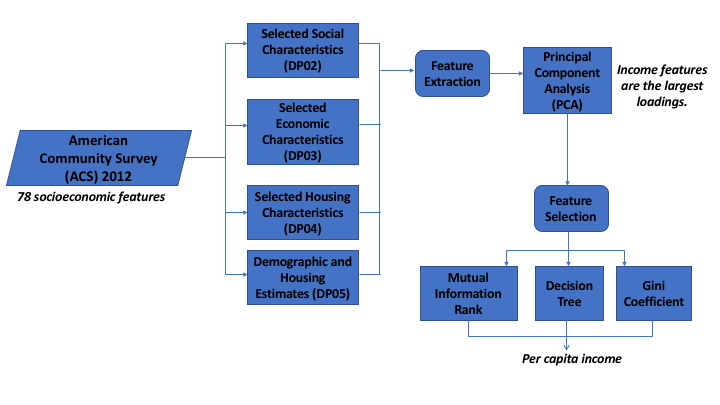}
	\caption{\textbf{Feature collection, extraction, and selection}. We collect 78 socioeconomic features from the ACS data. Based on PCA, the largest loadings correspond to income features (11 variables). Different feature selections (Mutual Information Rank, Decision Tree and Gini Coefficient) follow after. Per capita income stands out comparing to the rest of features.} 
\label{fig:A3a}       
\end{figure*}

\clearpage

\begin{figure*}[!htb]
\centering
\includegraphics[width=\textwidth]{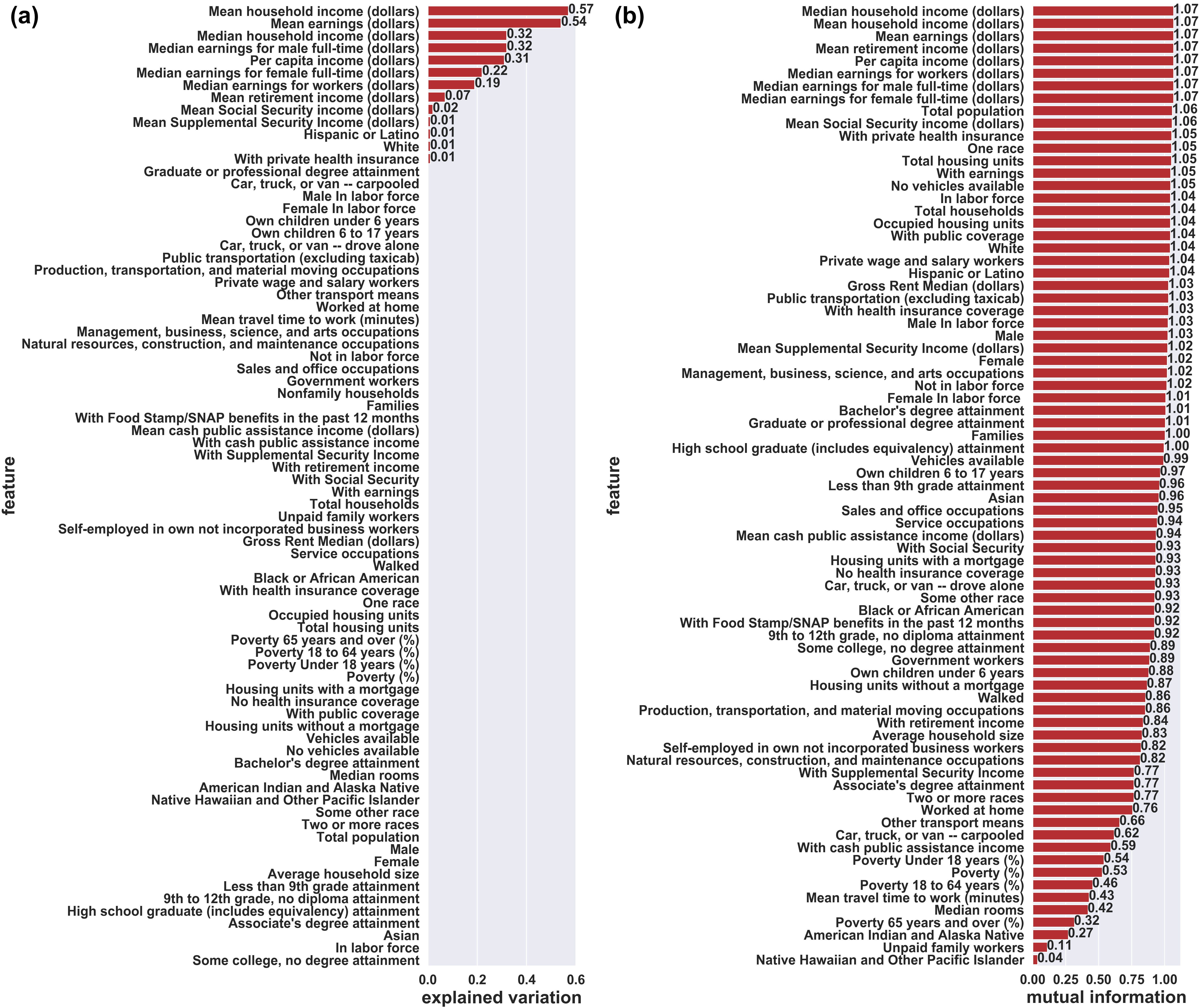}
\caption{\textbf{Principal Component Analysis (PCA) and Mutual Info (MI) Rank}. To determine the fittest feature in reflecting socioeconomic aspect, number of feature  extraction and selection techniques are applied accordingly. (a) We take the first principal component which explains 93.5 \% of the variability. (b) We estimate MI for discrete target variable (SES) to measure the dependency between variables. MI with higher value indicates higher dependency, thus, more explanatory contributions to SES are gained. PCA and MI Rank show corresponding results where variables associated with incomes appear on the top list.} 
\label{fig:A3b}
\end{figure*}

The result from feature selection in Fig. \ref{fig:A3b} confirms the previous finding given by feature extraction. Mutual Information (MI) Rank suggests that income features have the highest contribution to mutual information of SES. In Decision Tree, per capita income remains the most important to SES. Gini Coefficient indicates that mean cash public assistance (0.46) has the highest sensitivity in capturing inequality, followed by per capita income (0.40). 

We choose per capita income over mean cash public assistance due to several reasons. Firstly, the coverage of mean cash public assistance is limited to particular segment in society. Secondly, it will be an issue when we are going further for a comparative study across countries since not every country has similar cash public assistance program. Last, per capita income reflects various sources of income, including salary, social assistance, and wealth. Therefore, it is considered as a comprehensive measurement for average income and widely available.

\section{Matrix measures}
\label{SM B}
\subsection{Empirical stratification matrices}
\label{SM B1}

\begin{figure*}[!ht]
 \centering
    \includegraphics[width=0.7\linewidth]{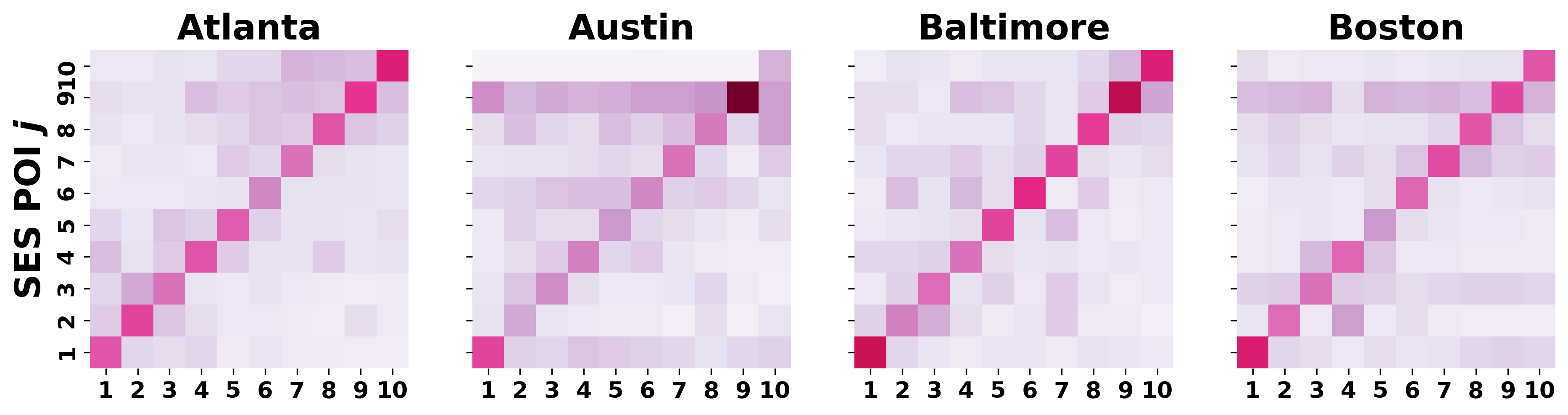}
    \includegraphics[width=0.7\linewidth]{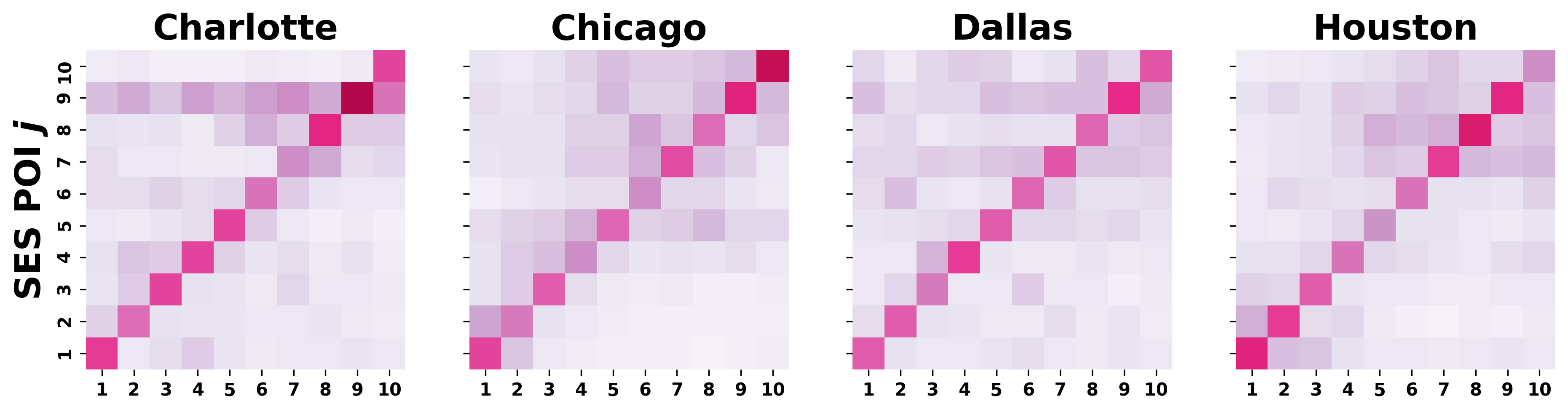}
    \includegraphics[width=0.7\linewidth]{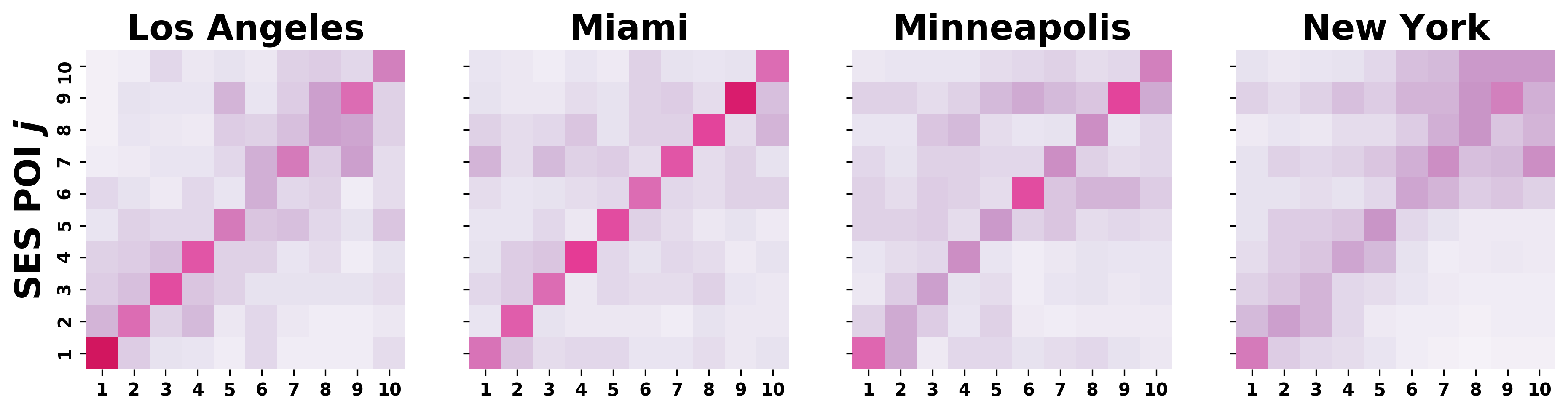}
    \includegraphics[width=0.7\linewidth]{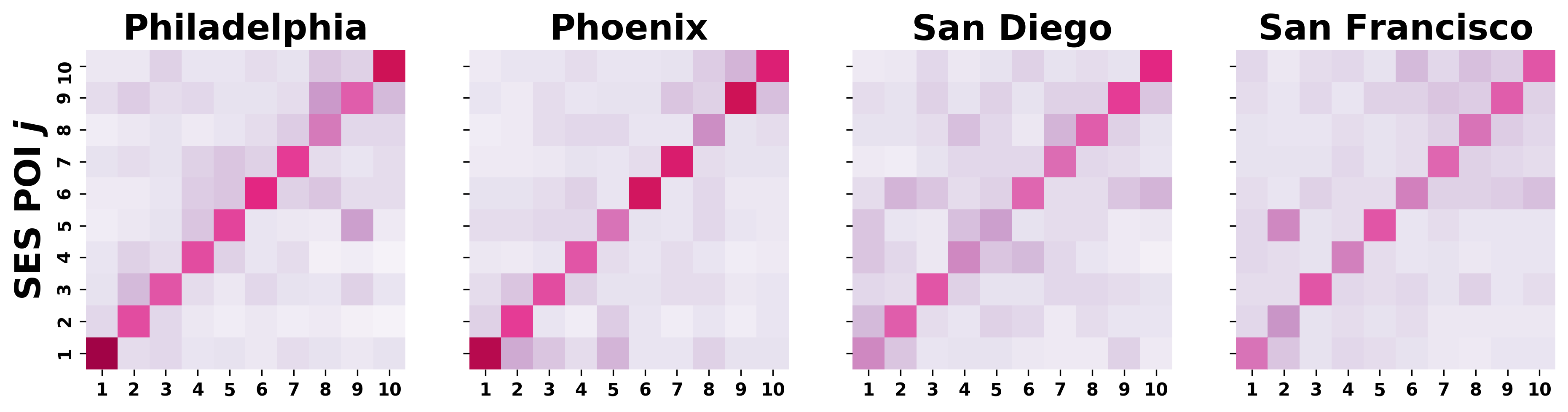}
    \includegraphics[width=0.7\linewidth]{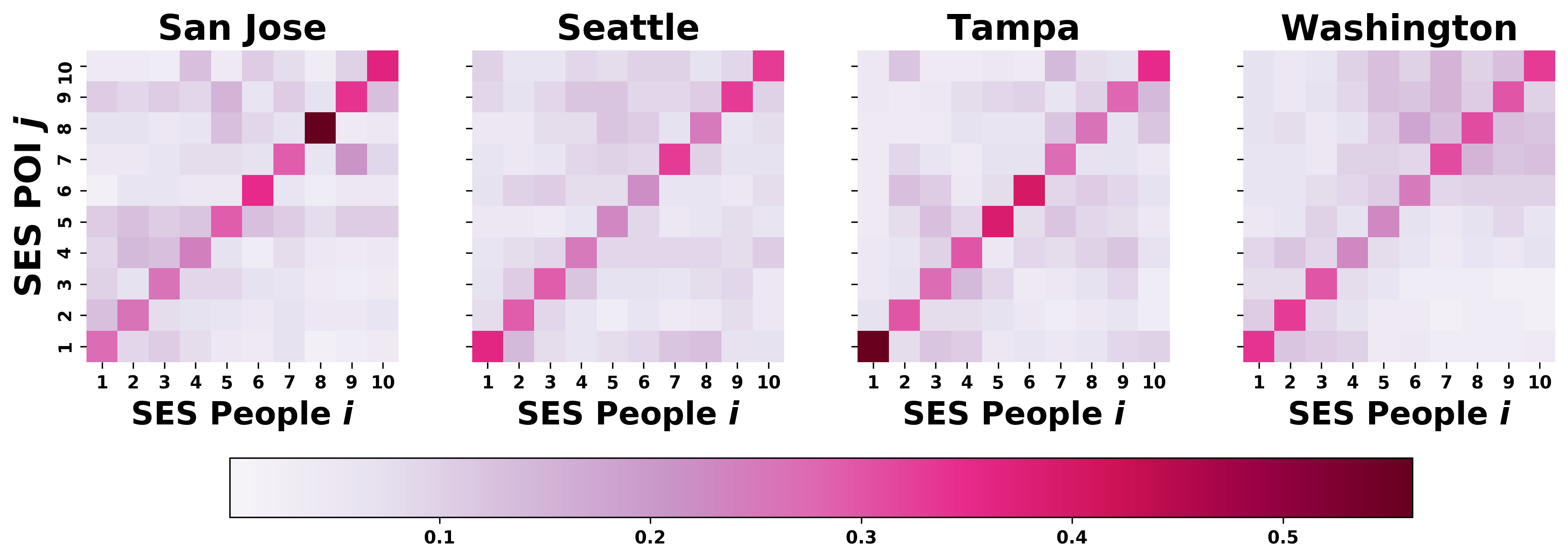}
	\caption{\textbf{Empirical stratification matrices $M_{i,j}$}. Each bin represents the visiting probabilities of individuals from a given class to places of different classes. The darker colour shades of bins represent larger visiting probability.}
\label{fig:B1}       
\end{figure*}

Mixing pattern in mobility network conceptualised in Section 3.1. measures the proportion of visits by people to places stratified by their respective SES. According to Empirical Stratification Matrices $M_{i,j}$ shown in Fig. \ref{fig:B1}, we find that individual mobility in some cities are less stratified than others in which colour of bins fades away from diagonal elements such as in New York and Austin. In contrast, strong empirical homophily mixing appears in most cities, for instance in Phoenix, Philadelphia, and Tampa. 

\clearpage

\subsection{Normalised stratification matrices}
\label{SM B2}

\begin{figure*}[!ht]
 \centering
    \includegraphics[width=0.7\linewidth]{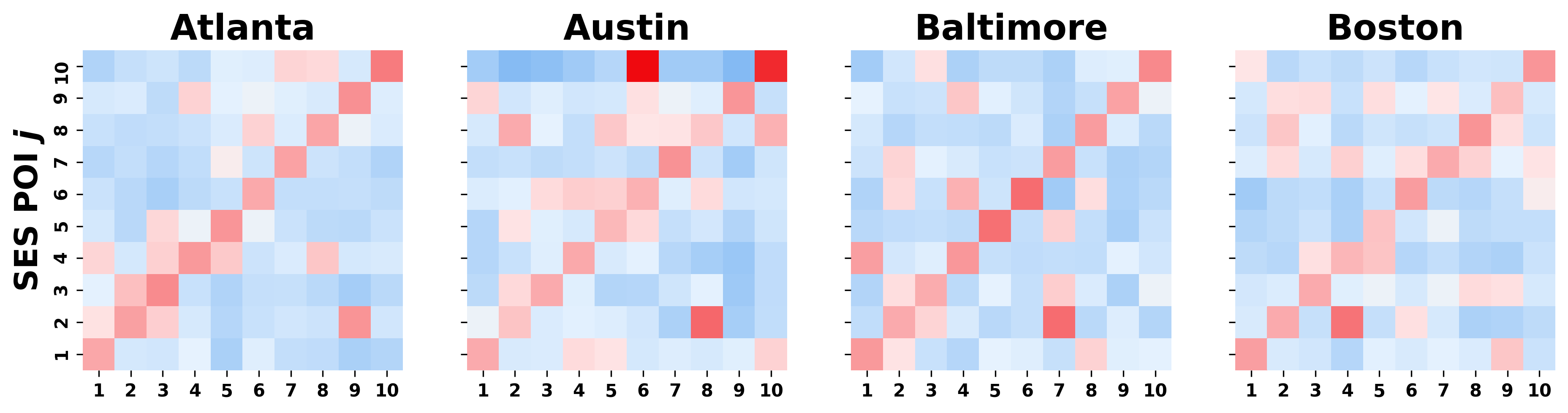}
    \includegraphics[width=0.7\linewidth]{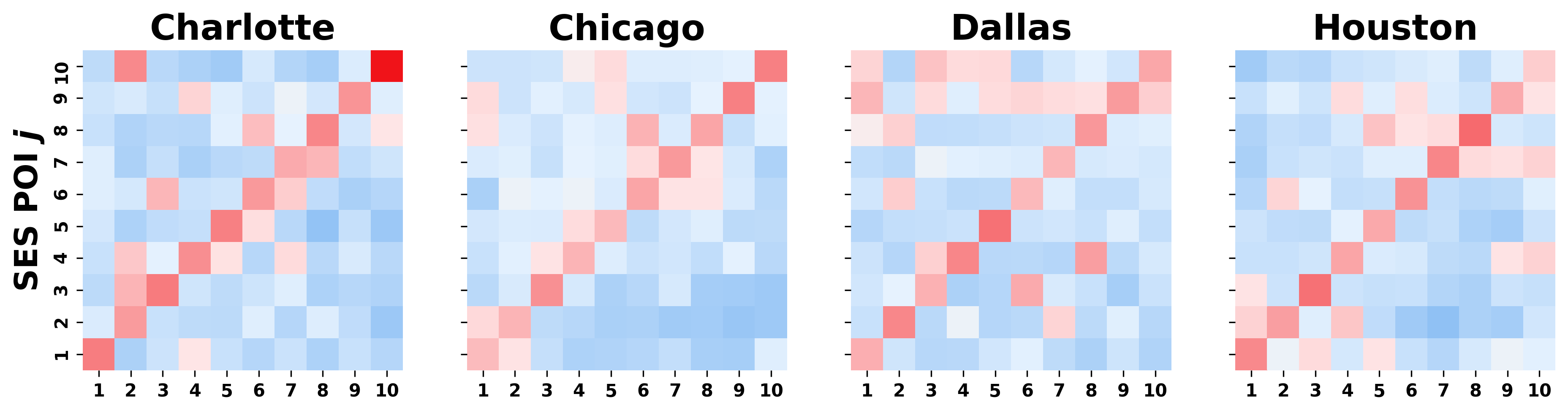}
    \includegraphics[width=0.7\linewidth]{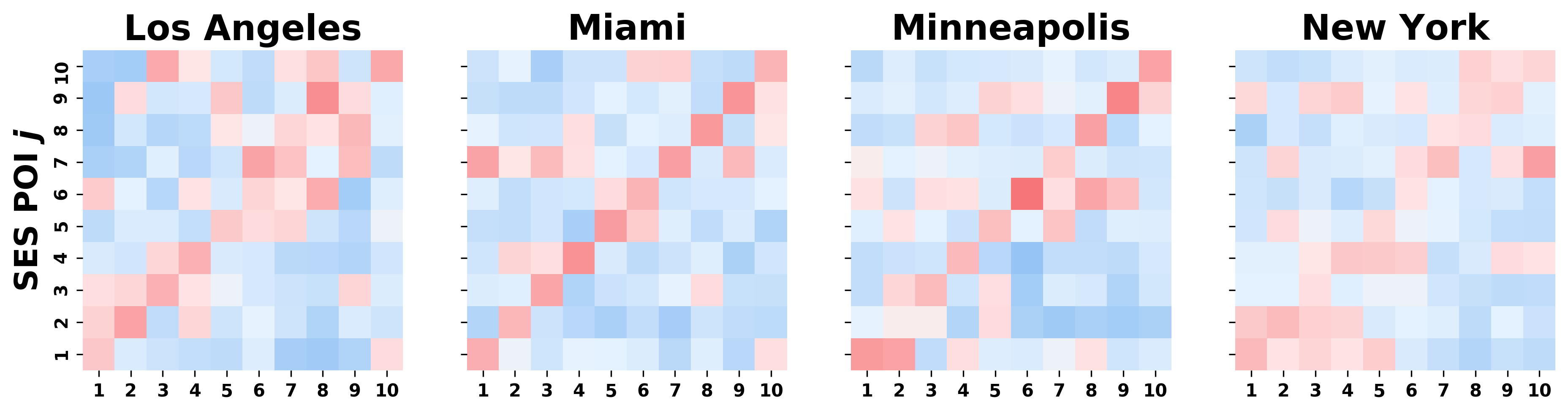}
    \includegraphics[width=0.7\linewidth]{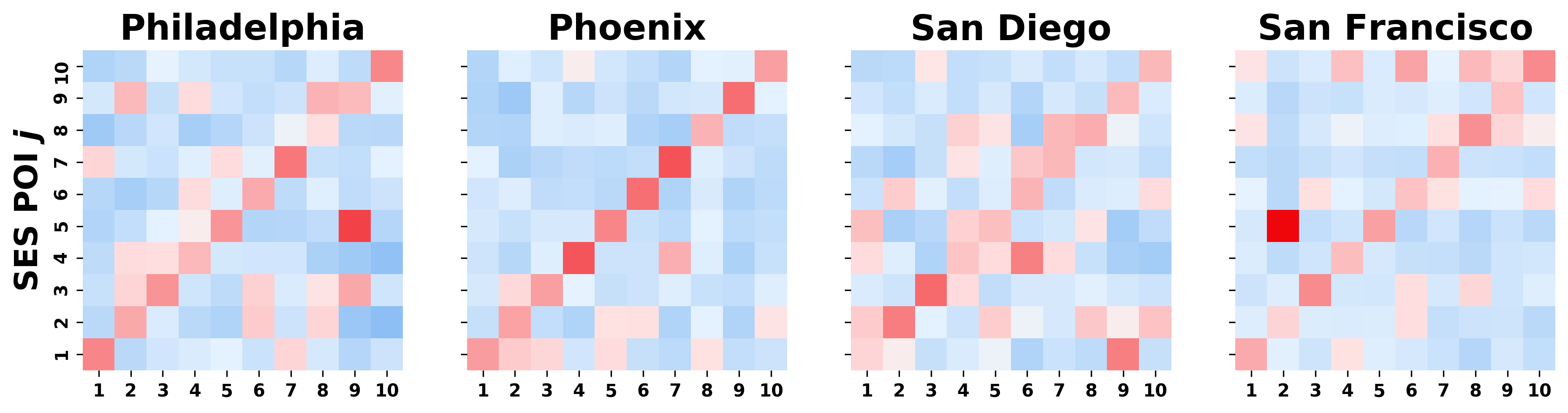}
    \includegraphics[width=0.7\linewidth]{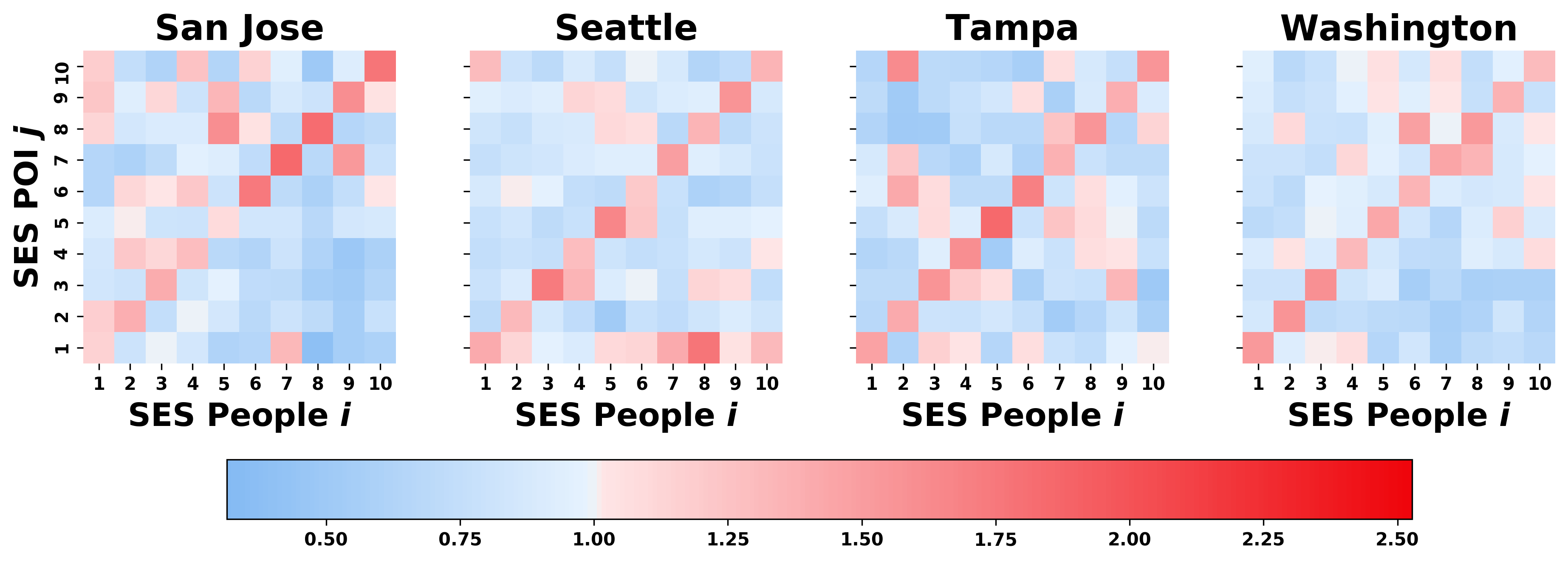}
	\caption{\textbf{Normalised stratification matrices $N_{i,j}$}. Each bin contains the fraction of the empirical and randomised stratification matrices. Colour gradients ranges from darker blue to darker red, taking the condition of lower to higher frequency visit than expected.} 
\label{fig:B2}       
\end{figure*}

We replicate methodological approach of Empirical Stratification Matrices $M_{i,j}$ used in Section 3.1 to construct Normalised Stratification Matrices $N_{i,j}$. In Fig. \ref{fig:B2}, red colour amplifies the higher frequency visit than expected, while blue denotes fairly less visit. Furthermore, the presence of red blocks in the upper diagonal elements signifies the upwards bias where people are aspired to drop by at places with higher SES. Bold red gradient along the diagonal exhibits homophily mixing. 

\clearpage

\section{Bias measures}
\label{SM C}
\subsection{Individual bias z-score}
\label{SM C1}

\begin{figure*}[!ht]
 \centering
    \includegraphics[width=0.75\linewidth]{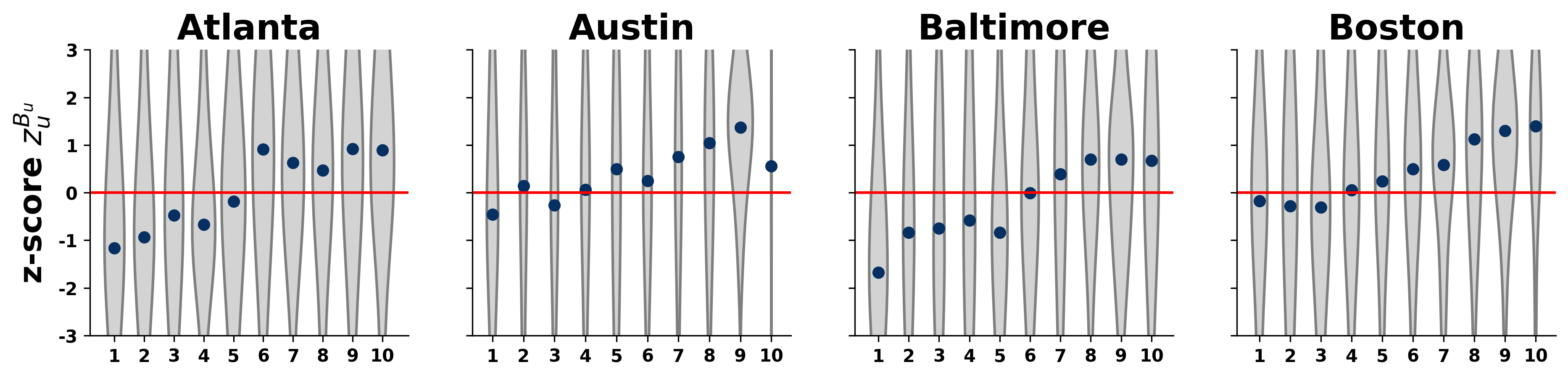}
    \includegraphics[width=0.75\linewidth]{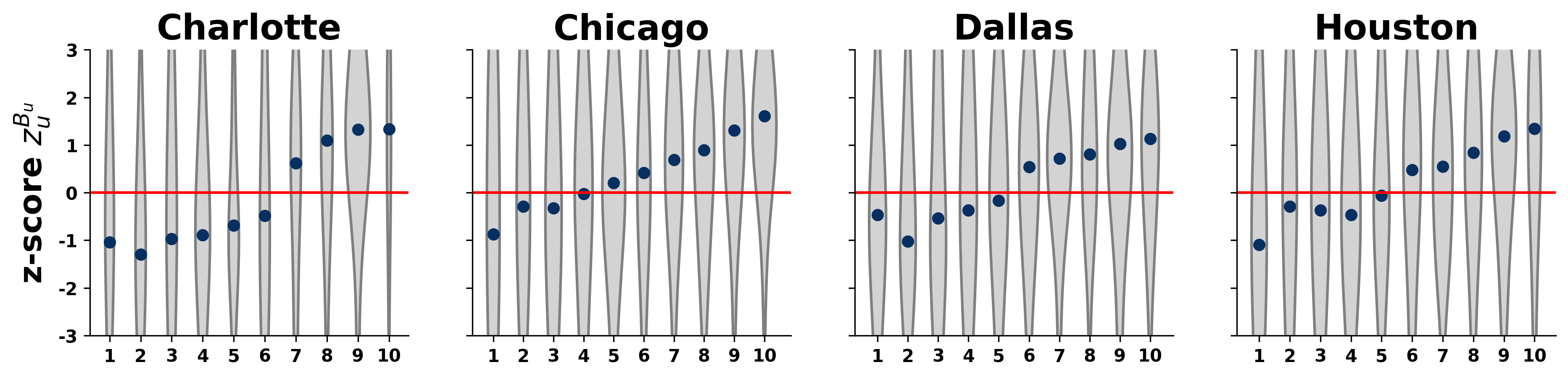}
    \includegraphics[width=0.75\linewidth]{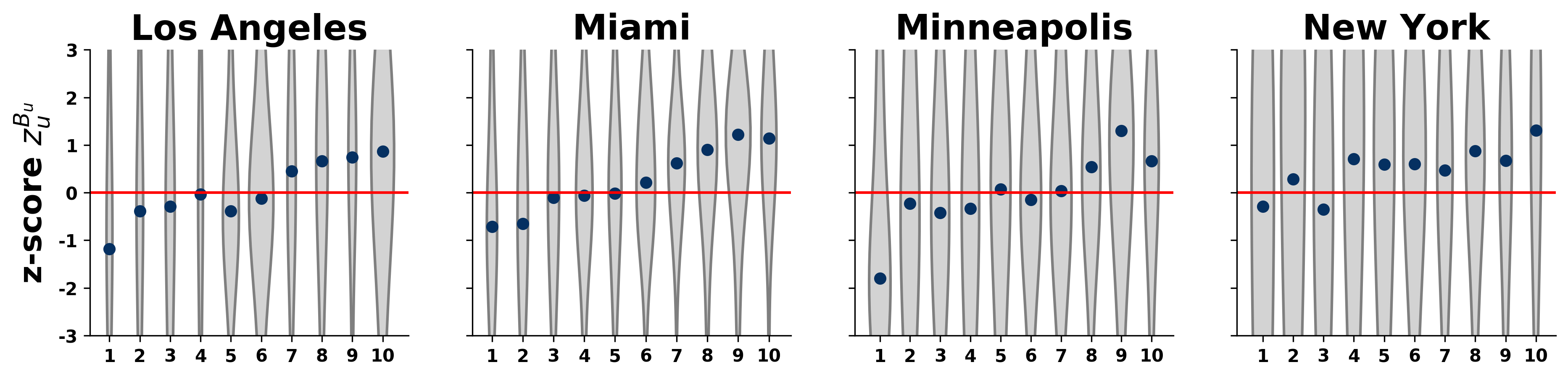}
    \includegraphics[width=0.75\linewidth]{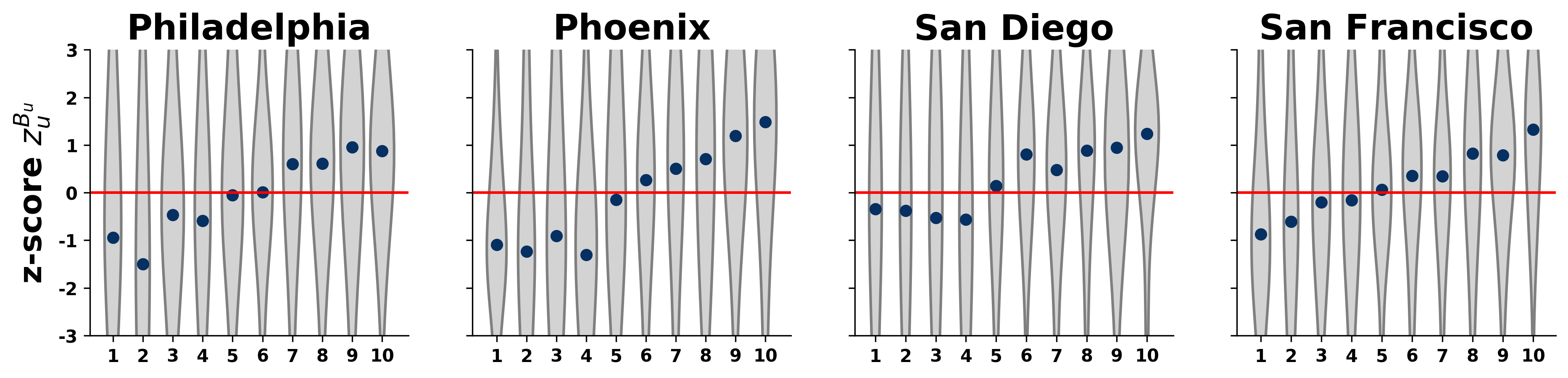}
    \includegraphics[width=0.75\linewidth]{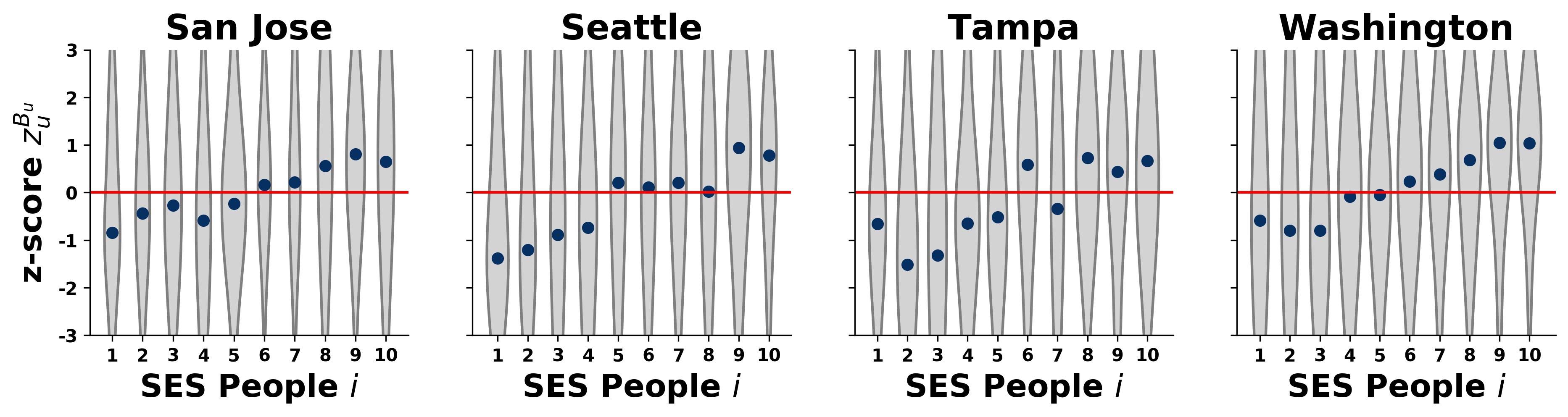}
	\caption{\textbf{Individual Bias z-Score $z_u^{B_u}$}. Class distributions of individual z-scores reflects how much the individual bias differs from the expected bias for an individual who chooses places to visit with the same frequency as before but selects them from a given set of places dictated by others within the same socioeconomic class.} 
\label{fig:C1}       
\end{figure*}

Individual Bias z-score $z_u^{B_u}$ measures how far empirical individual bias score deviates from the random mobility model. Technical formulation is explained in Section 3.2. Similar to typical individual mobility in other studies, individual mobility in our data is also characterised by large jumps between two types of people: small proportion of people with large trajectory and largely dominant type with small trajectory. Fig. \ref{fig:C1} reveals the coexistence of upward and downward biases in terms of visiting patterns to other socioeconomic classes in any class.

\clearpage

\subsection{Class-level bias z-score}
\label{SM C2}
\begin{figure*}[!ht]
 \centering
    \includegraphics[width=0.75\linewidth]{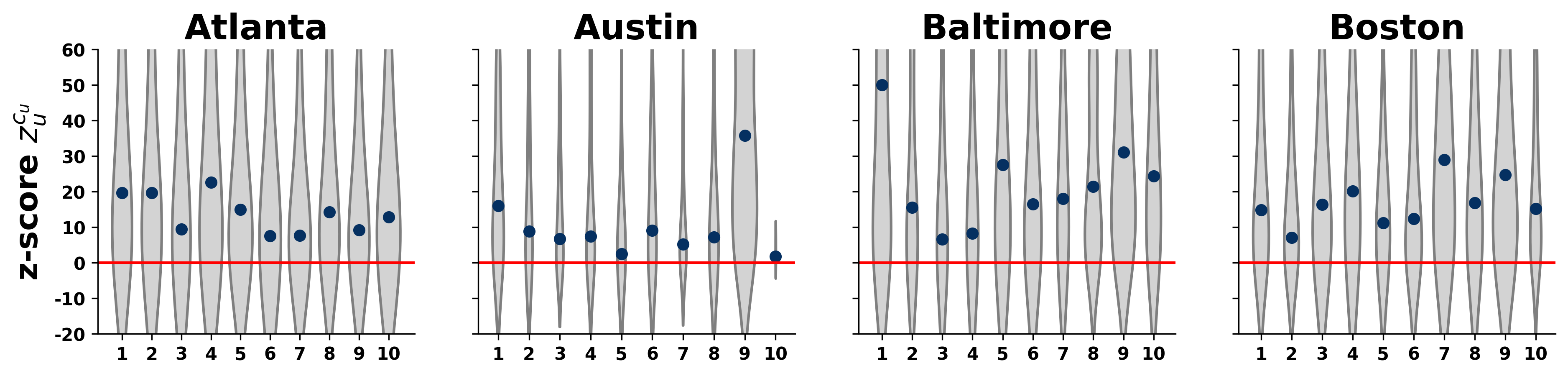}
    \includegraphics[width=0.75\linewidth]{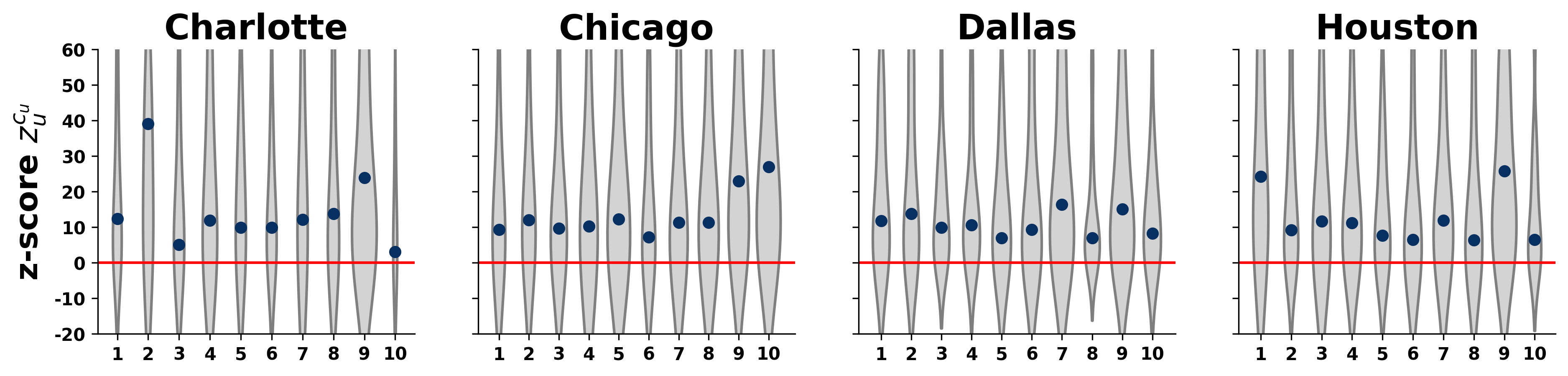}
    \includegraphics[width=0.75\linewidth]{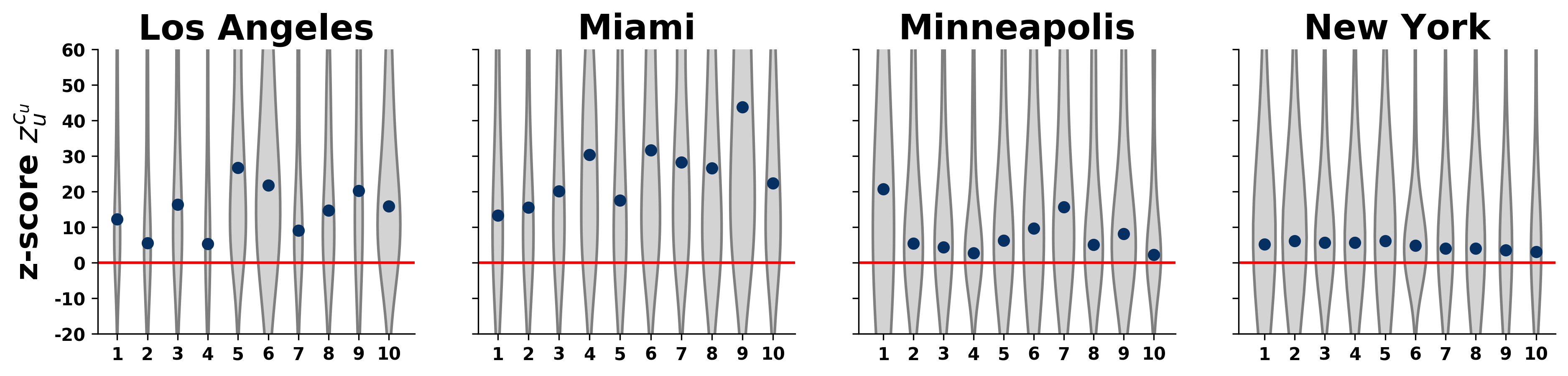}
    \includegraphics[width=0.75\linewidth]{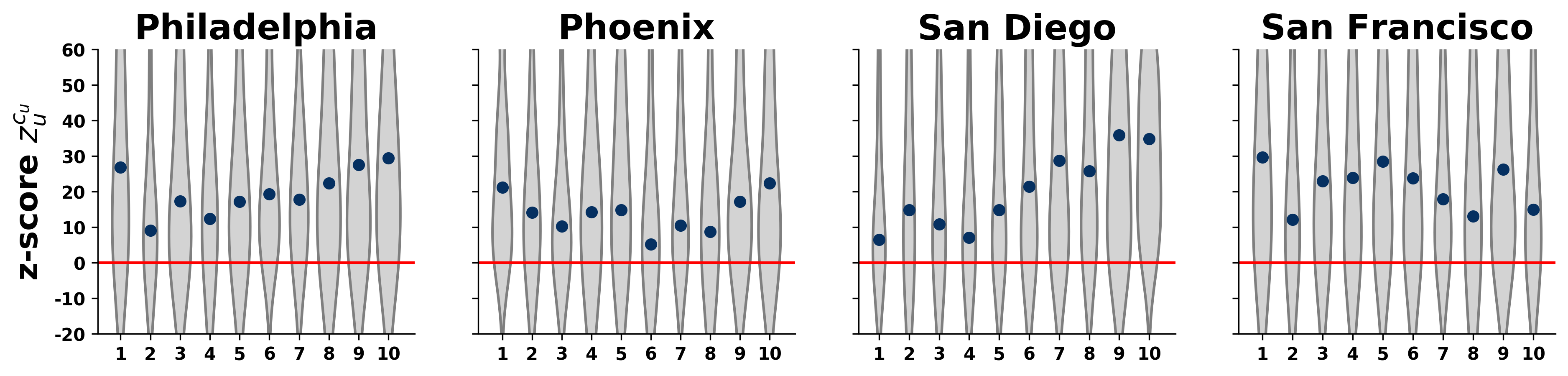}
    \includegraphics[width=0.75\linewidth]{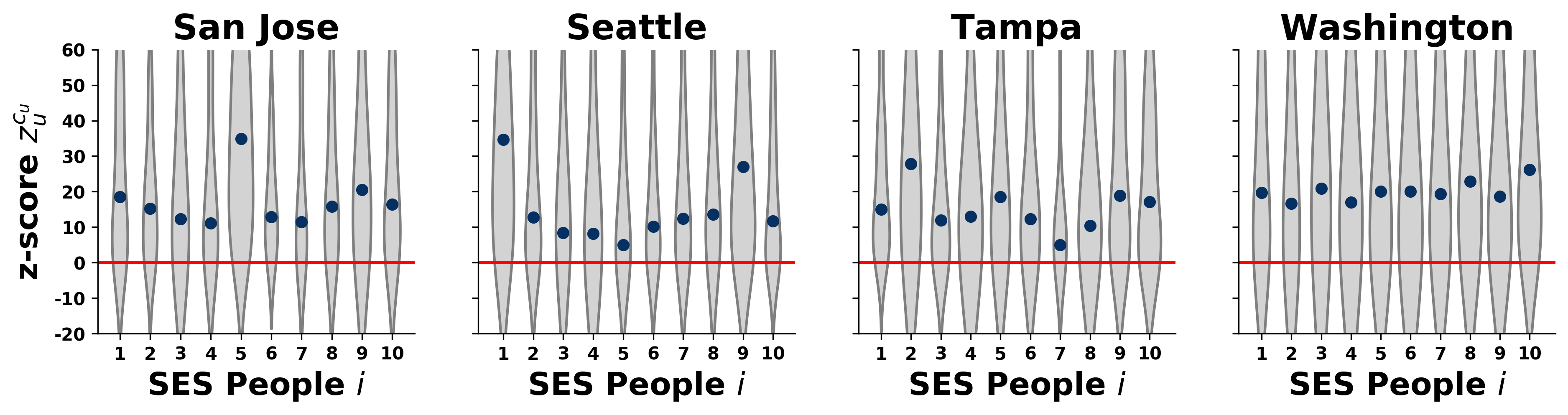}
	\caption{\textbf{Class-level Bias z-score $z_u^{c_u}$}. Distribution of class-level biased z-scores reflects directly how much the individual behaviour deviates from the expected level, when the individual could choose randomly places to visit from a given set dictated by others from the same socioeconomic class.} 
\label{fig:C2}       
\end{figure*}

Class-level Bias z-score $z_u^{c_u}$ (Fig. \ref{fig:C2}) is formalised to quantify the difference between typical visited places in individual trajectory from the trajectory in random class average. Composition of visit pattern differs with respect to SES. Many places, usually within own range, are visited many times along the trajectory, meanwhile, some others are only visited away less frequent. We dedicate Section 3.2 to formulate and discuss the measurement. 

\clearpage

\section{Out-of-class matrix measures}
\label{SM D}
\subsection{Out-of-class empirical stratification matrices}
\label{SM D1}

\begin{figure*}[!ht]
 \centering
    \includegraphics[width=0.7\linewidth]{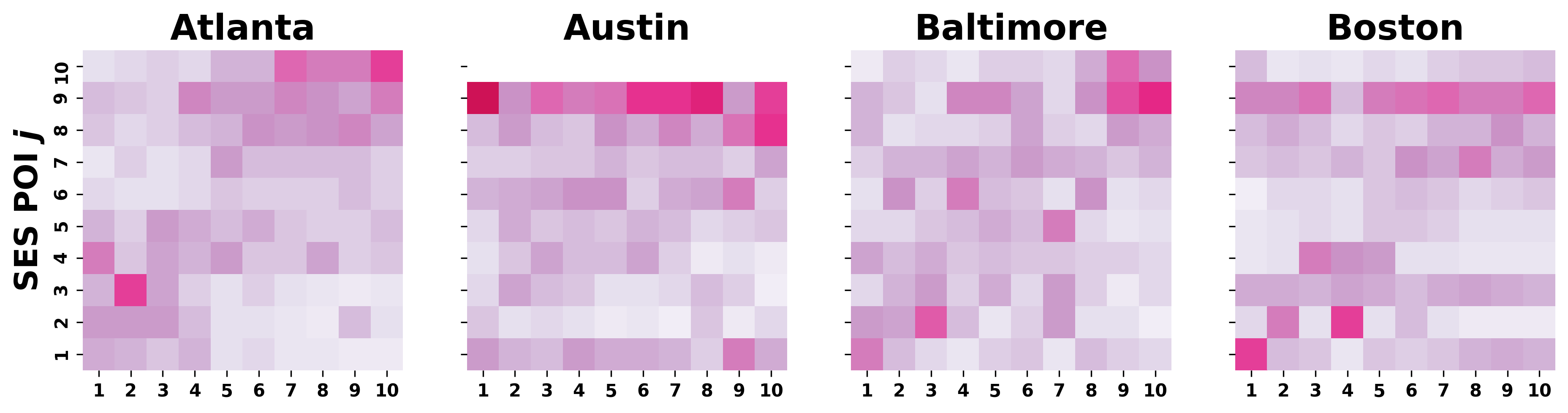}
    \includegraphics[width=0.7\linewidth]{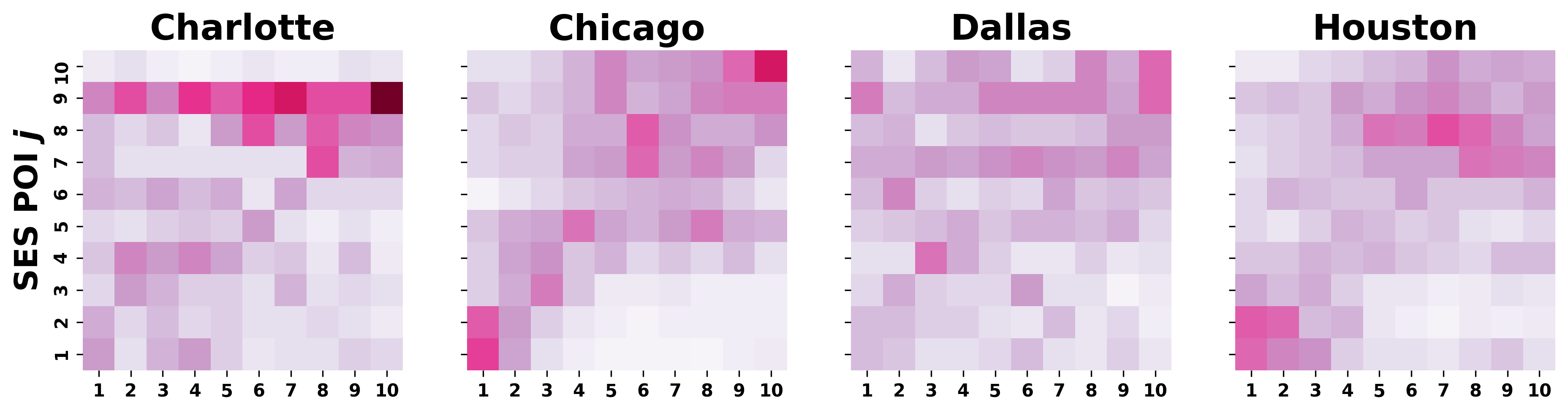}
    \includegraphics[width=0.7\linewidth]{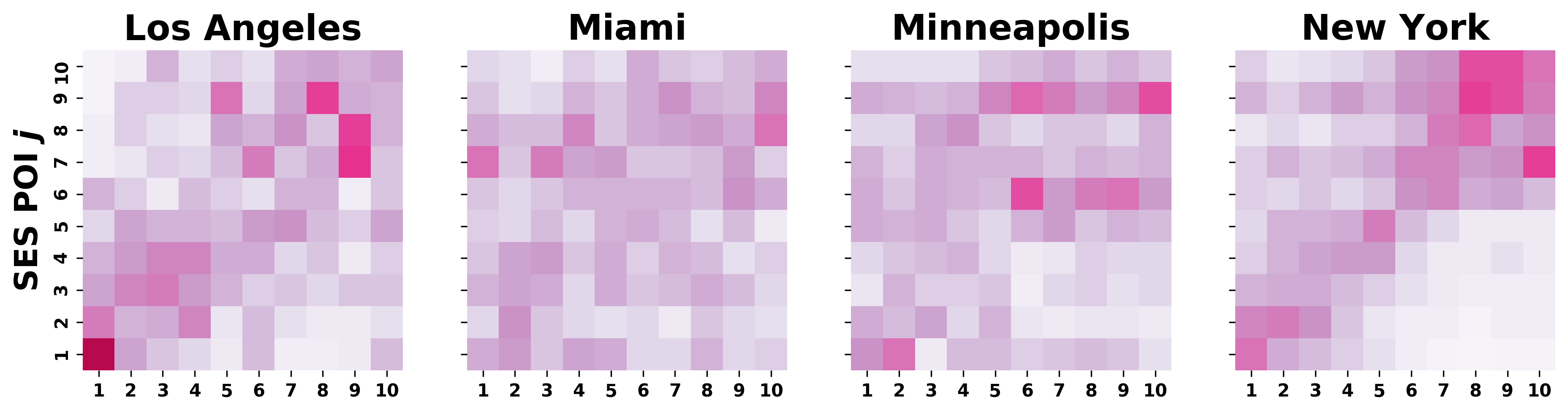}
    \includegraphics[width=0.7\linewidth]{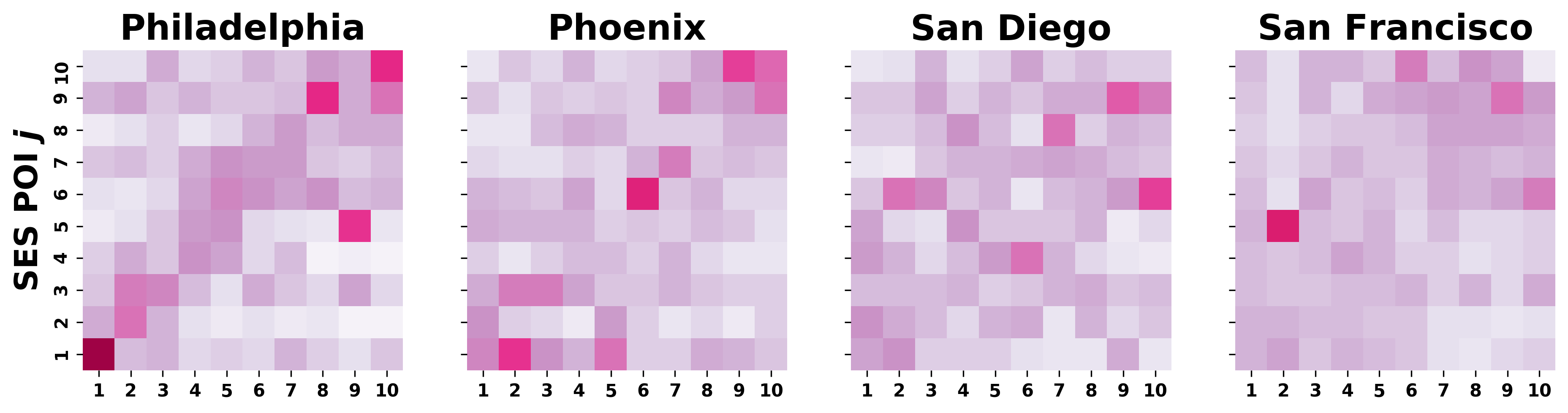}
    \includegraphics[width=0.7\linewidth]{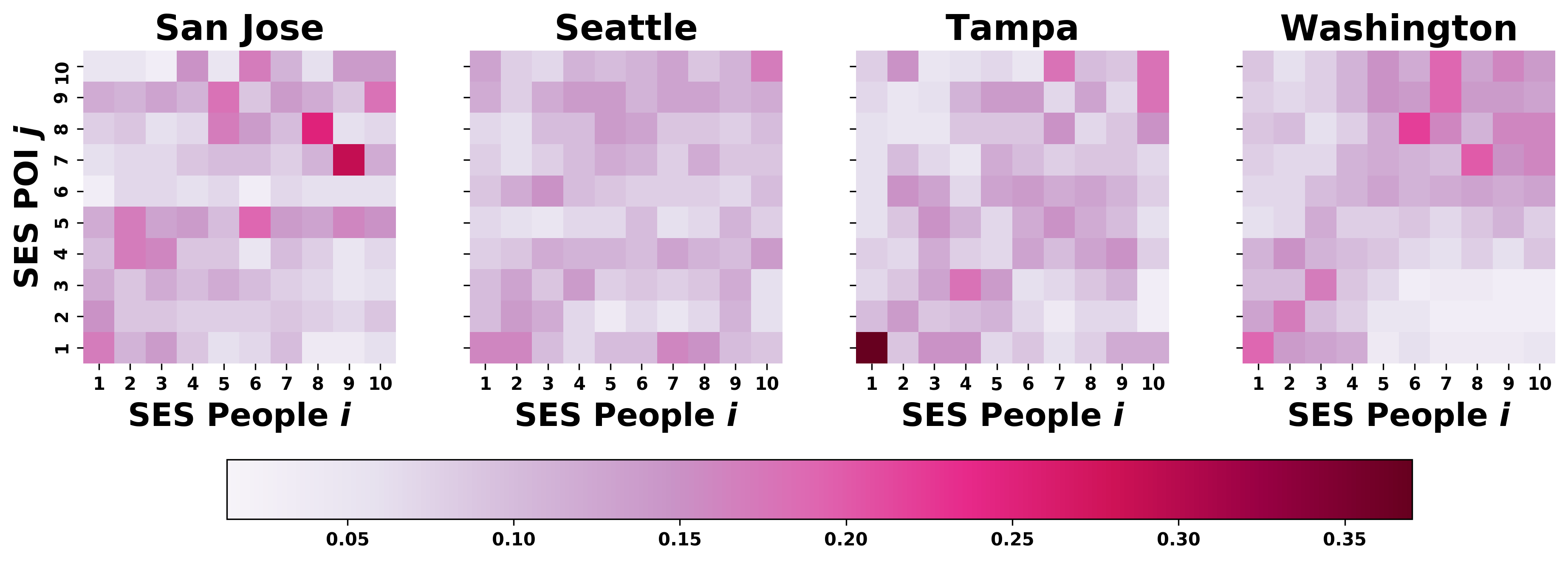}
	\caption{\textbf{Out-of-class Empirical Stratification Matrices $Mc_{i,j}$}. Probabilities that individuals from a given class visit to places of different classes are visualised with colour shades. The darker the bin, the higher visiting probability.}
\label{fig:D1}       
\end{figure*}

Out-of-class Empirical Stratification Matrices $Mc_{i,j}$ shown in Fig. \ref{fig:D1} takes the same methodological approach as Empirical Stratification Matrices $M_{i,j}$ shown in Fig. \ref{fig:B1}. The two differs in term of census tract scope where out-of-class excludes own census tract in each individual trajectory. This step is taken into account to control distance effect that contributes to homophily mixing condition. Even though we observe less diagonality here, some stratification patterns remain in several cities.

\clearpage

\subsection{Out-of-class normalised stratification matrices}
\label{SM D2}

\begin{figure*}[!ht]
 \centering
    \includegraphics[width=0.7\linewidth]{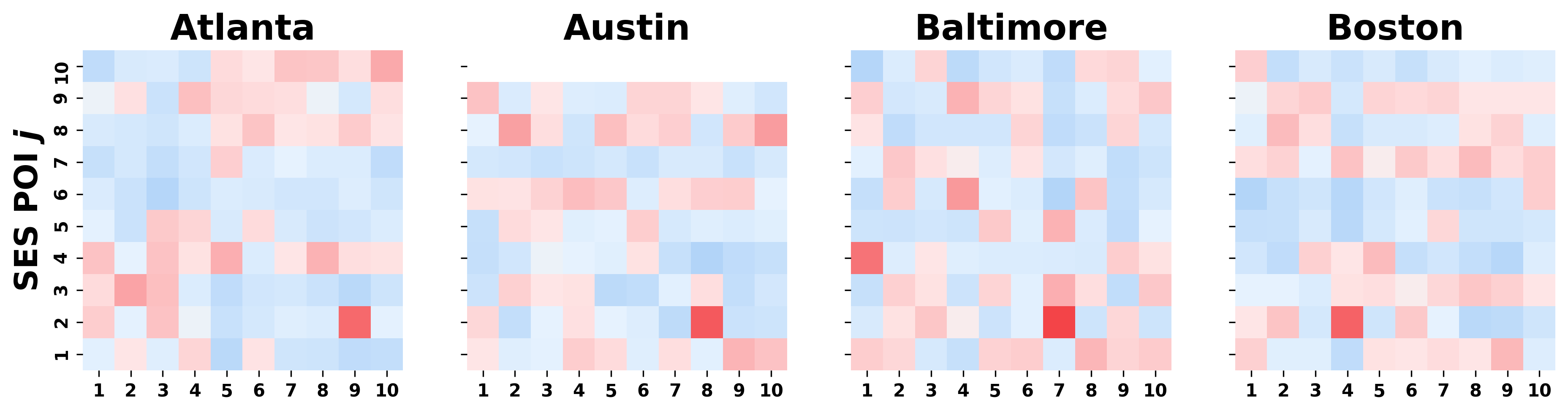}
    \includegraphics[width=0.7\linewidth]{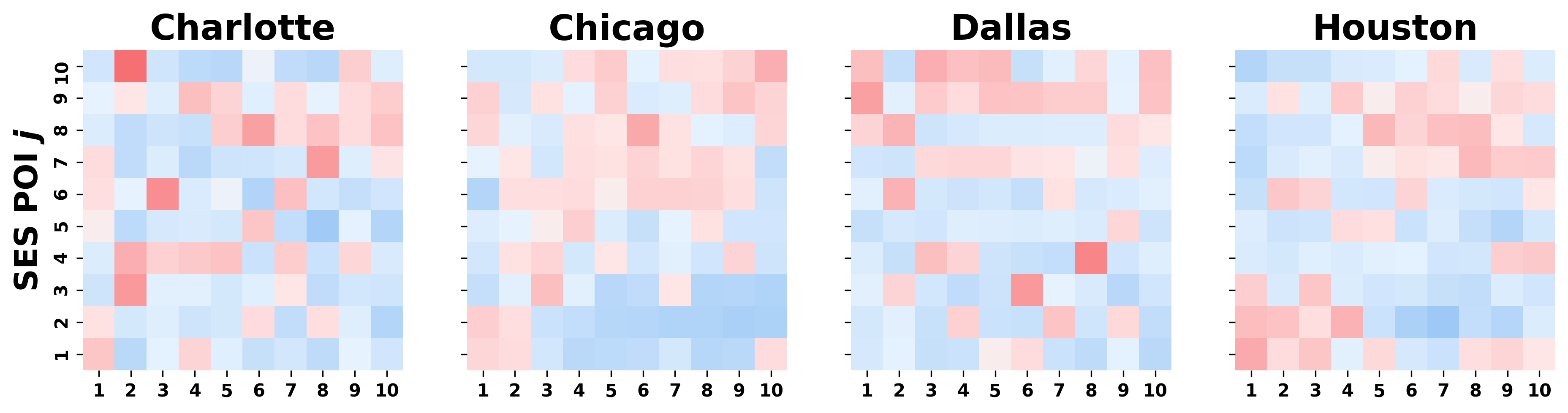}
    \includegraphics[width=0.7\linewidth]{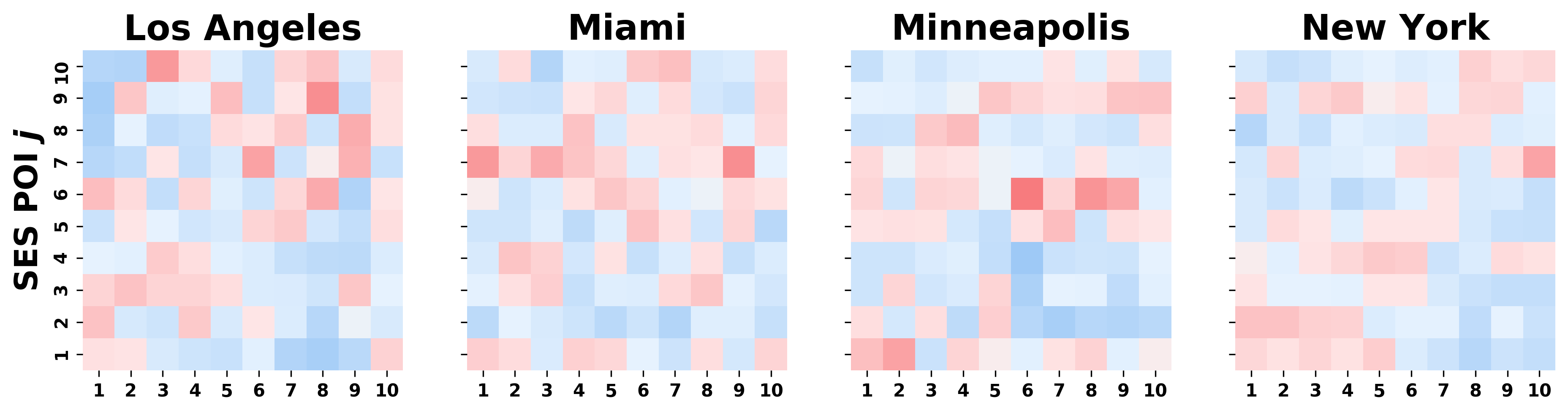}
    \includegraphics[width=0.7\linewidth]{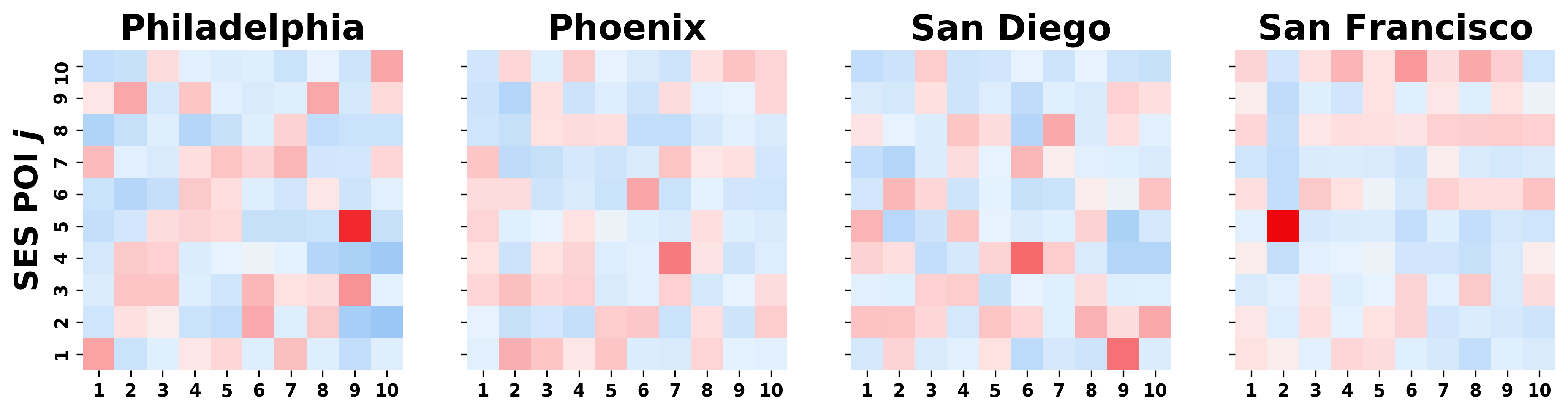}
    \includegraphics[width=0.7\linewidth]{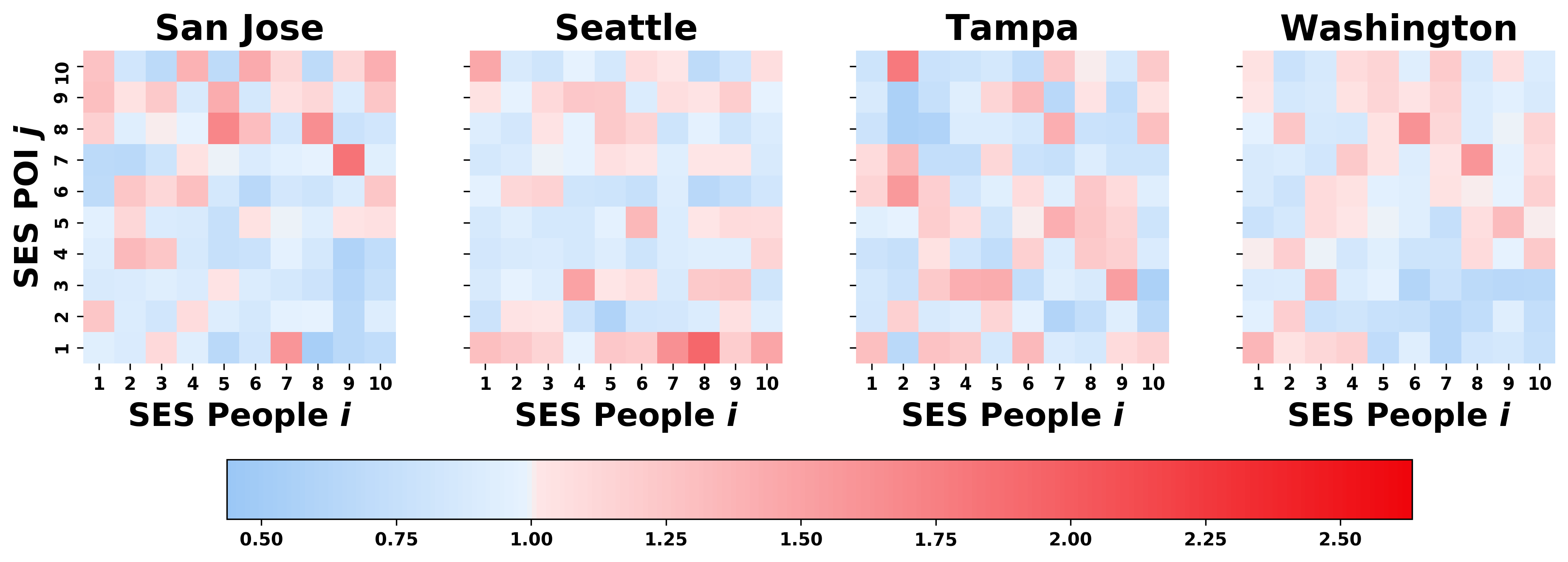}
	\caption{\textbf{Out-of-class Normalised Stratification Matrices $Nc_{i,j}$}. Defined as the fraction of the empirical and randomised stratification matrices without own census tract, out-of-class normalised stratification matrices reveals the visiting patterns of upward bias (red gradient), downward bias (blue gradient), and no bias (white) tendency across a SES pair of people and visited places in their trajectories.}
\label{fig:D2}       
\end{figure*}

We construct Out-of-class Normalised Stratification Matrices $Nc_{i,j}$ (Fig. \ref{fig:D2}) in order to examine the robustness of Out-of-class Empirical Stratification Matrices $Mc_{i,j}$ (Fig. \ref{fig:D1}). Normalised version of out-of-class stratification matrices is generated to measure the difference in magnitude the out-of-class empirical stratification matrices make in comparison to random visit occurrence. We observe that upward bias tendency (red gradient in upper diagonal matrices) is still considerably present in some cities and such distinct pattern doesn't emerge by chance.

\section{Out-of-class bias measures}
\label{SM E}
\subsection{Out-of-class individual bias z-score}
\label{SM E1}

\begin{figure*}[!htb]
 \centering
    \includegraphics[width=0.75\linewidth]{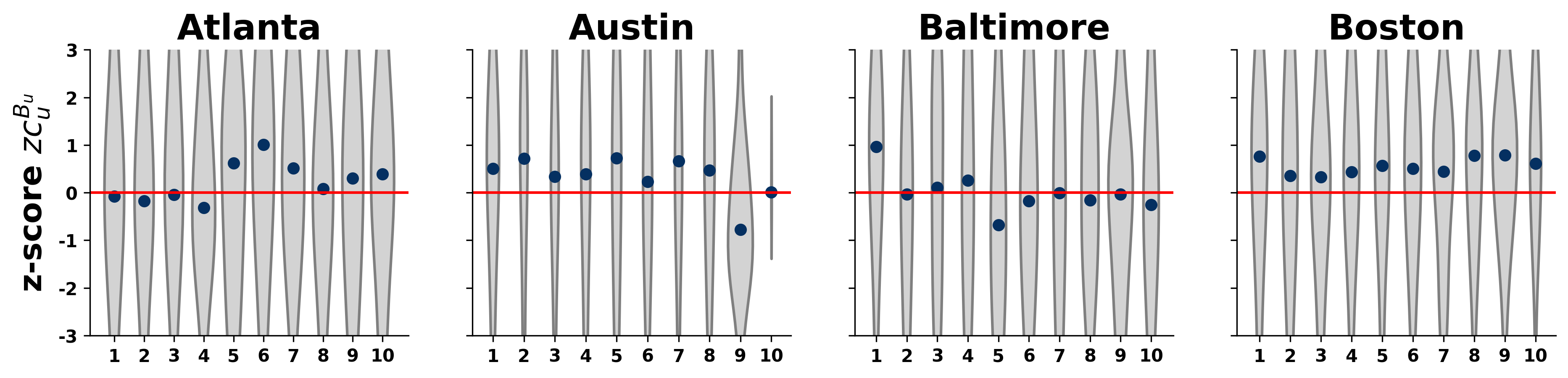}
    \includegraphics[width=0.75\linewidth]{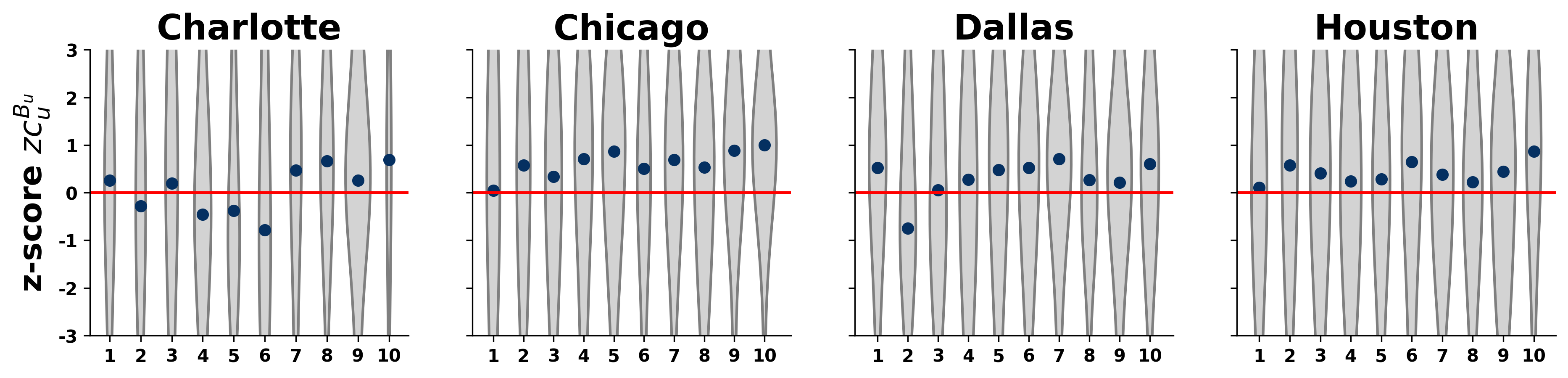}
    \includegraphics[width=0.75\linewidth]{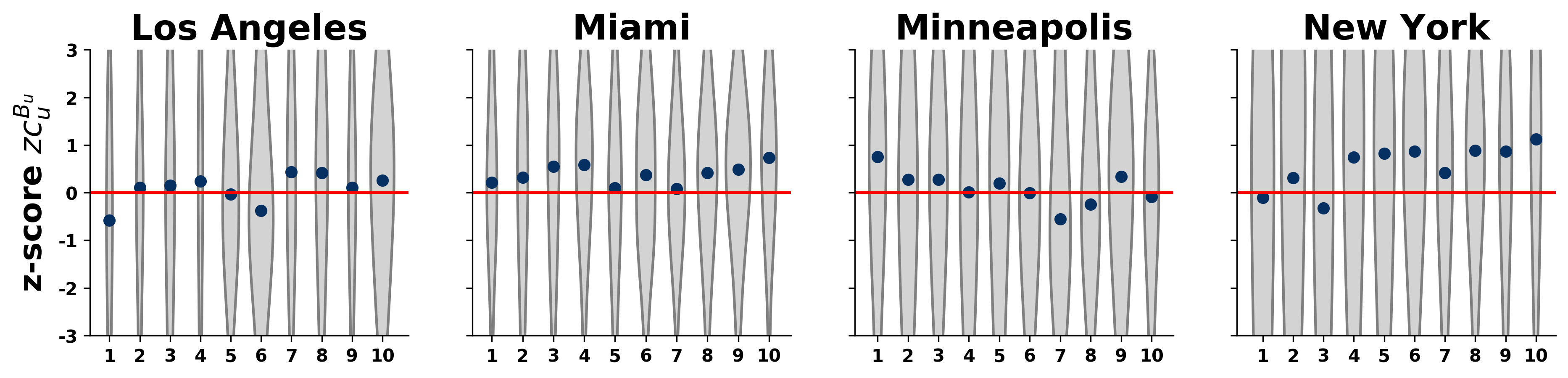}
    \includegraphics[width=0.75\linewidth]{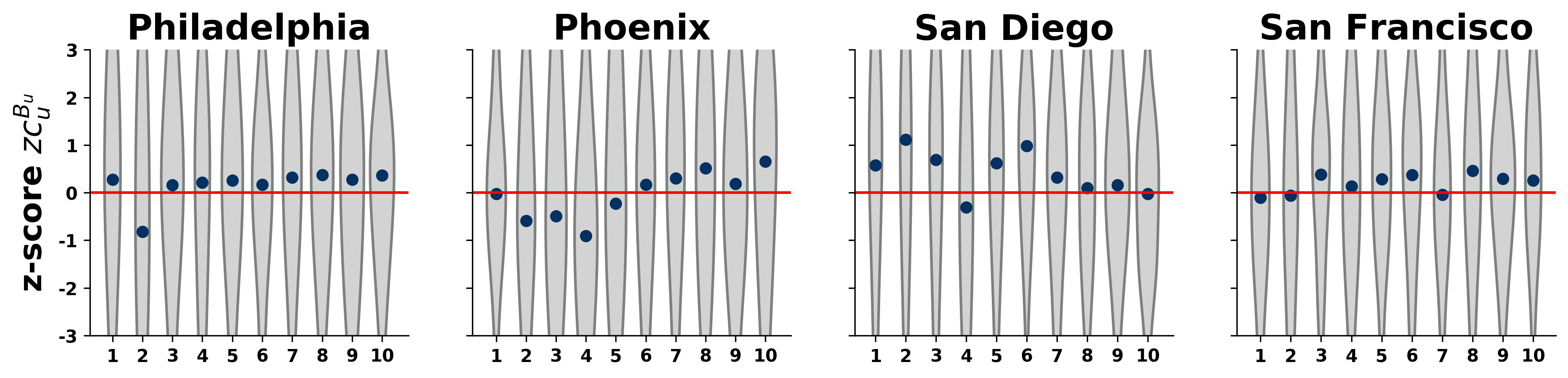}
    \includegraphics[width=0.75\linewidth]{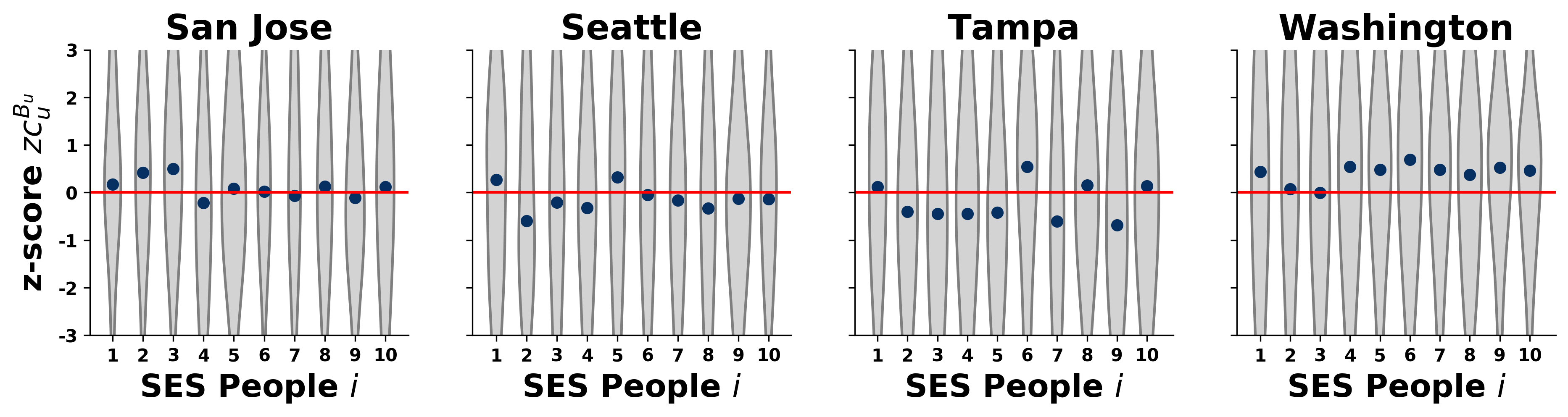}
	\caption{\textbf{Out-of-class Individual Bias z-score $zc_u^{B_u}$}. We show class level distributions and their median values (blue dots) for each socioeconomic class after removing own census tract. Unbiased condition is depicted by horizontal red line. Upper red line areas contain upward biases in terms of visiting patterns, in contrast to downward biases on the lower part.} 
\label{fig:E1}       
\end{figure*}

Out-of-class Individual Bias z-score $zc_u^{B_u}$ depicted in Fig. \ref{fig:E1} takes the same methodological approach as Individual Bias z-score $z_u^{B_u}$ shown in Fig. \ref{fig:C1}. The two differs in term of census tract scope where out-of-class excludes own census tract in each individual trajectory. This step is taken into account to control distance effect that contributes to biases in visiting patterns, recalling that mobility mostly takes place in home census tract or nearby locations. 

\clearpage

\subsection{Out-of-class class-level bias z-score}
\label{SM E2}

Out-of-class class-level bias z-score $zc_u^{B_u}$ indicates if this individual bias is weaker or stronger than expected from random behaviour, considering the absence of visit to places located in own census tract. It reflects directly how much the individual behaviour deviates from the expected level, when the individual could choose randomly places to visit from a given set dictated by others from the same socioeconomic class. 

\begin{figure*}[!htb]
 \centering
    \includegraphics[width=0.75\linewidth]{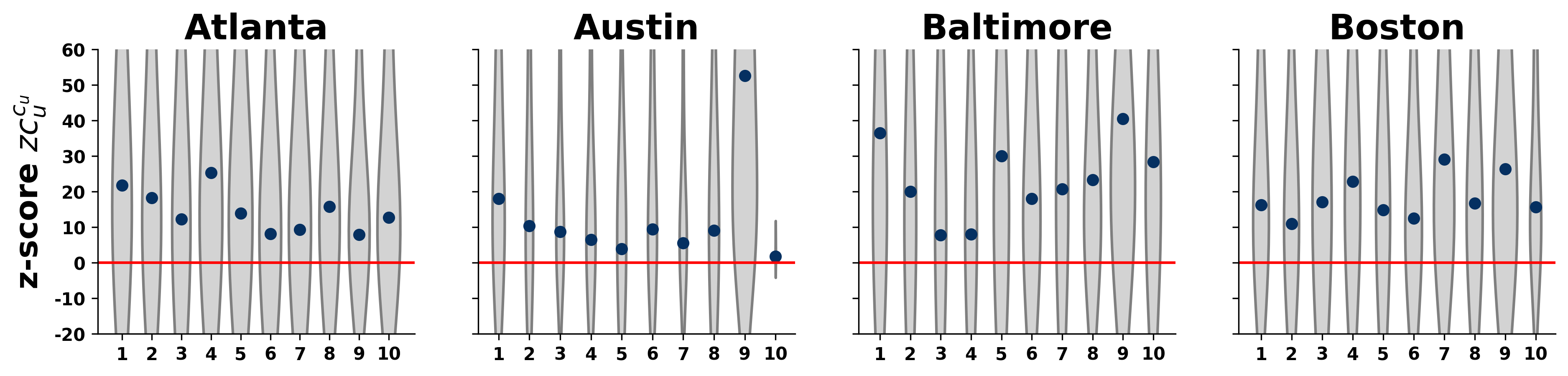}
    \includegraphics[width=0.75\linewidth]{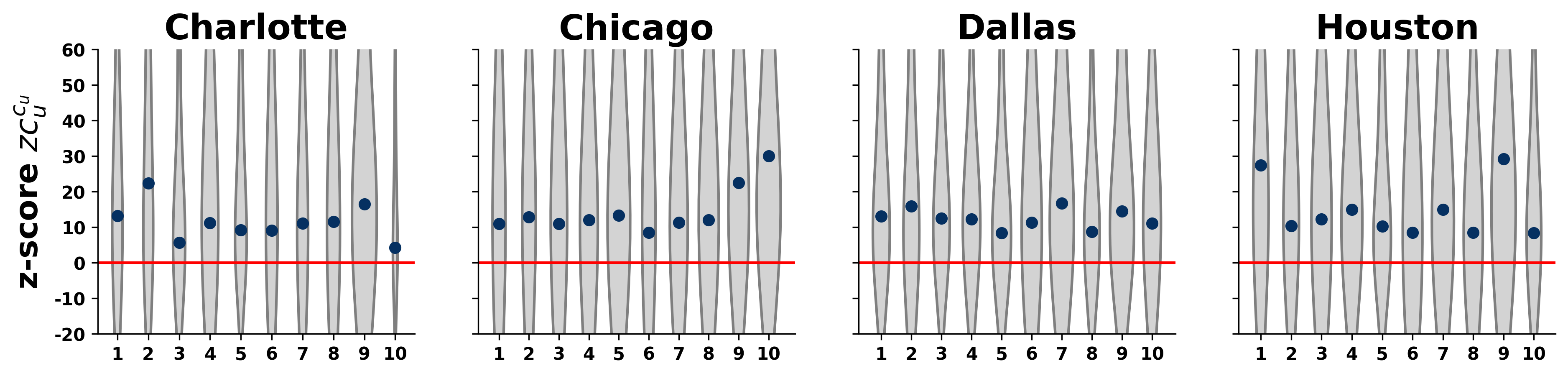}
    \includegraphics[width=0.75\linewidth]{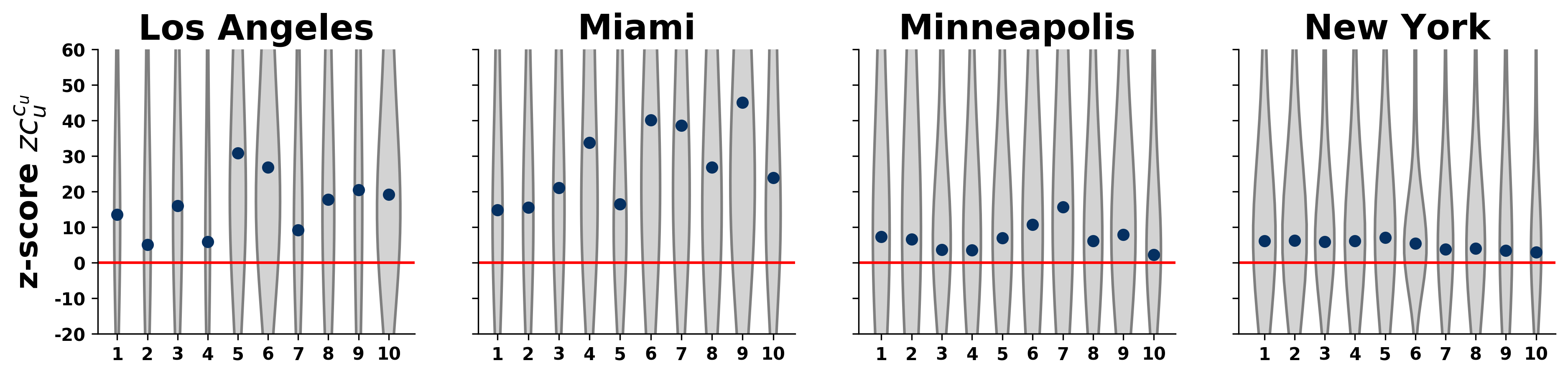}
    \includegraphics[width=0.75\linewidth]{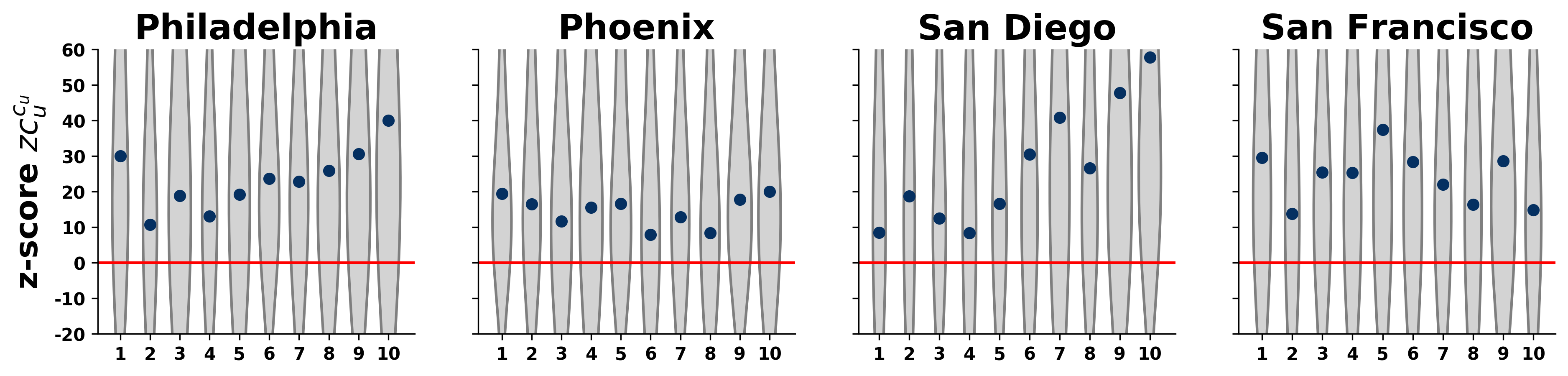}
    \includegraphics[width=0.75\linewidth]{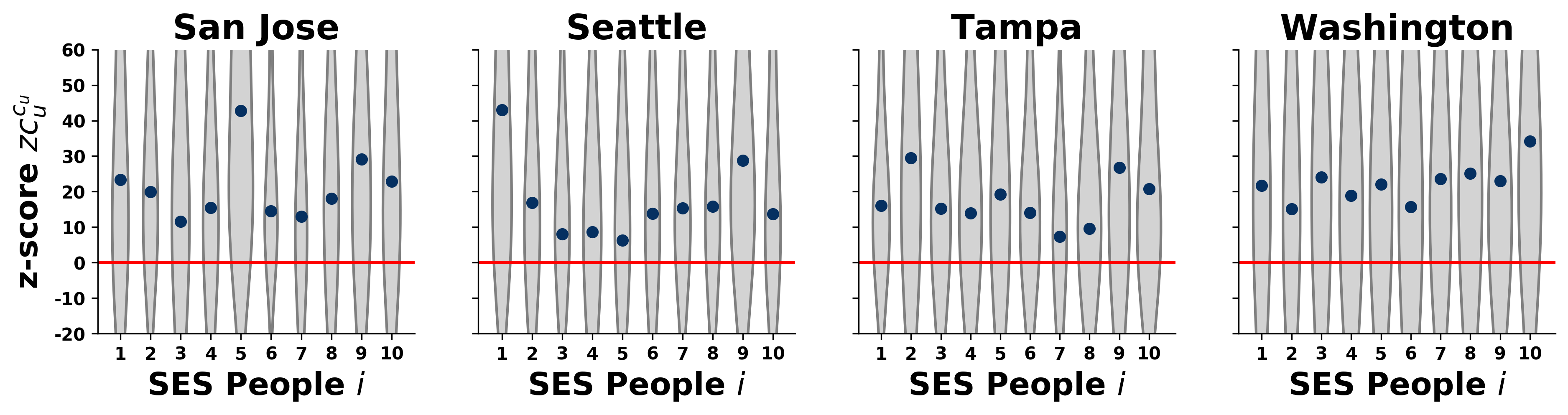}
	\caption{\textbf{Out-of-class Class-level Bias z-Score $zc_{u}^{c_u}$}. After removing own census tract, distributions are shown for each socioeconomic class with their median values as blue points. Above the red unbiased median line, an upward visiting bias holds, otherwise upward visiting bias is prevalent.} 
\label{fig:E2}       
\end{figure*}

\clearpage

\section{Out-of-class measures in Houston, New York, and San Francisco}
\label{SM F}

\begin{figure*}[ht!]
    \centering
        \includegraphics[width=0.75\linewidth]{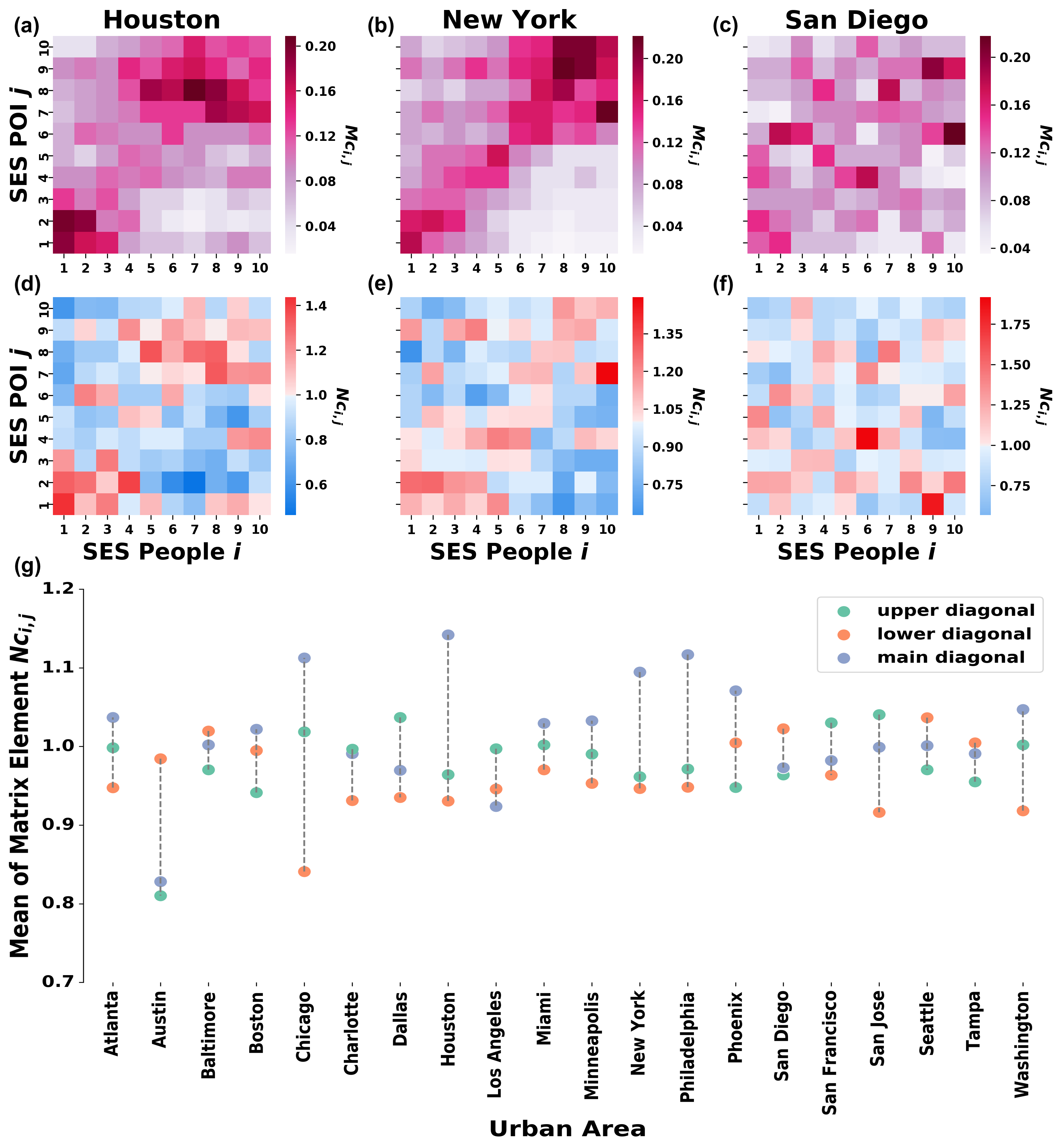}
        \caption{\textbf{Out-of-class socioeconomic stratification matrices.} Out-of-class socioeconomic stratification matrices are constructed based on individual trajectories or a set of locations visited by each individual after the removal of own census tract. 
        (a) The out-of-class empirical stratification matrices $Mc_{i,j}$, showing the probabilities that individuals from a given class visit to places of different classes. The darker colour shades of bins represent larger visiting probability. (b) The out-of-class normalised stratification matrices $Nc_{i,j}$, defined as the fraction of the empirical and randomised stratification matrices without own census tract. We observe less diagonality in Houston (Fig. ~\ref{fig:F1}a and Fig. ~\ref{fig:F1}d), New York (Fig. ~\ref{fig:F1}b and Fig. ~\ref{fig:F1}e) and San Diego (Fig. ~\ref{fig:F1}c and Fig. ~\ref{fig:F1}f). Interestingly, upward bias tendency is still considerably present as seen in (Fig. ~\ref{fig:F1}g) where
        mean of upper diagonal matrix elements exceeds mean of lower diagonal matrix elements in 13 out of 20 urban areas, including Houston, New York, and San Diego.}
        \label{fig:F1}       
\end{figure*}

In Section ~\ref{SM D} and Section ~\ref{SM E}, we arise the issue of potential confounding factor driven by distance effect. It is important to realise that shortcoming may arise from this particular computation which disregards distance aspect. In dealing with that, we remove mobilities in own census tracts and recompute the stratification matrices as well as bias measures. In line with the discussion in the main text (Fig. 2), this section is dedicated to analyse the robustness and sensitivity of our homophily mixing and visiting bias measures in Houston, New York, and San Diego. We find that mobility is still stratified by SES as seen in Fig. \ref{fig:F1}a, \ref{fig:F1}b and \ref{fig:F1}c. Moreover, the presence of upward bias tendency is still visible in those cities by looking at Fig. \ref{fig:F1}d, \ref{fig:F1}e, and \ref{fig:F1}f. It is confirmed by the fact that mean of upper diagonal matrix elements surpass mean of lower diagonal matrix elements in \ref{fig:F1}g. 

\begin{figure*}[!htb]
    \centering
    \includegraphics[width=0.8\linewidth]{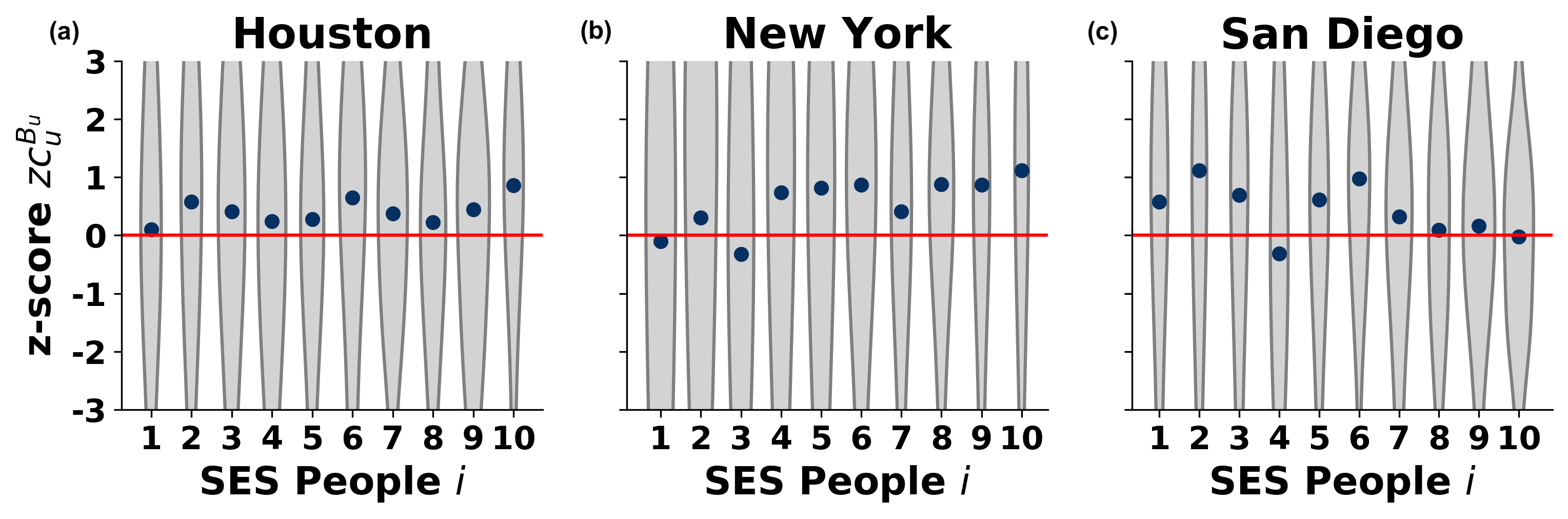}
    \caption{\textbf{Out-of-class Individual Bias z-score $zc_u^{B_u}$}. After removing own census tract, class level distributions and their median values are shown for each socioeconomic class. Horizontal red line at z-score=0 represents the condition of unbiasedness. Median values that appear on the top of red line suggests upward biases in terms of visiting patterns, otherwise downward biases persist. The blue dots show that while in general people from lower classes in New York (Fig. ~\ref{fig:F2}b) are less bias than expected, they could be as much biased regardless their socioeconomic classes in Houston (Fig. ~\ref{fig:F2}a) and San Diego (Fig. ~\ref{fig:F2}c).}
\label{fig:F2}
\end{figure*}

Taking a deeper analysis, we exploit the persistence of biased behaviour in mobility by employing procedural bias measures at individual and class level that have been introduced earlier in Section \ref{SM D} and \ref{SM E}. In the case of Out-of-class Individual Bias z-score $zc_u^{B_u}$, visiting bias is less expected among lower class people in New York (Fig. ~\ref{fig:F2}b), while anyone in Houston (Fig. ~\ref{fig:F2}a) and San Diego (Fig. ~\ref{fig:F2}c) can be as much biased in their visiting patterns  regardless their socioeconomic classes. Upward visiting bias is even stronger at class level in all three as none of z-score values fall below the red unbiased median line (Fig. ~\ref{fig:F2}a, Fig. ~\ref{fig:F2}b and Fig. ~\ref{fig:F2}c).

\begin{figure*}[!htb]
    \centering
        \centering
        \includegraphics[width=0.8\linewidth]{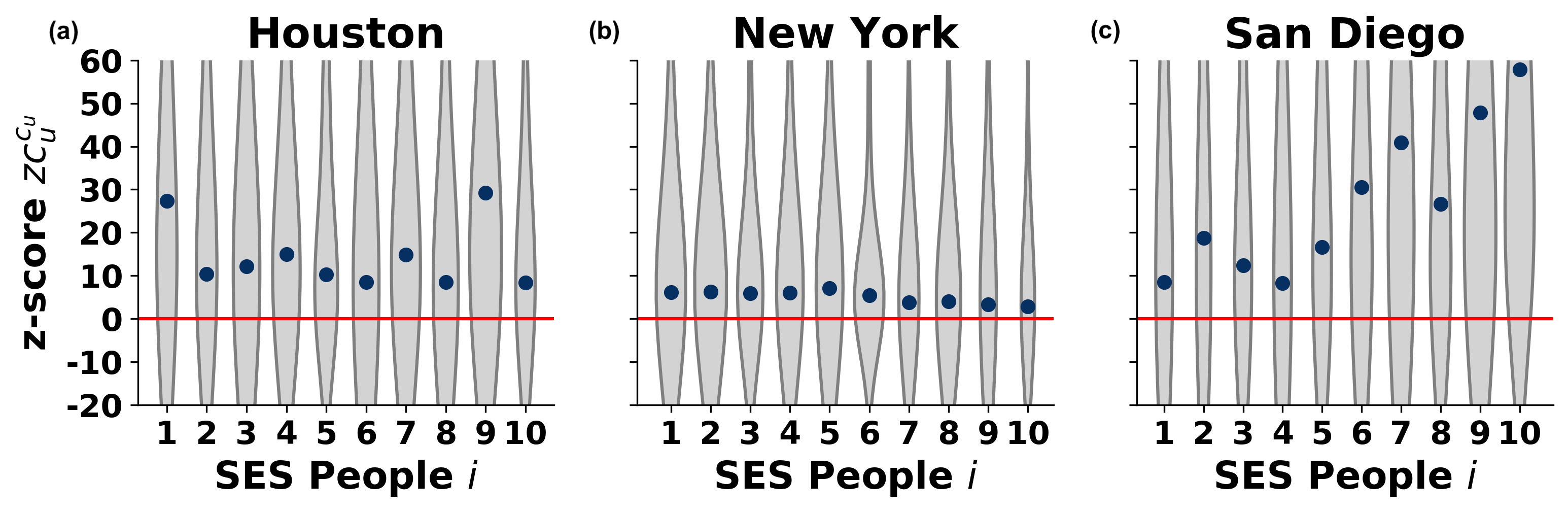}
        \caption{\textbf{Out-of-class Class-level Bias z-score $zc_{u}^{c_u}$}. Distribution of class-level biased z-scores as the function of socioeconomic classes. After removing own census tract, distributions are shown for each socioeconomic class with their median values as blue points for Houston (Fig.~\ref{fig:F3}a), New York (Fig.~\ref{fig:F3}b), and San Diego (Fig.~\ref{fig:F3}c). All z-score values remain above the red unbiased median line, therefore signalling an upward visiting bias.}
\label{fig:F3}       
\end{figure*}

\clearpage

\section{Homophily mixing measures}
\label{SM G}

We formulate \textbf{Dispersion Index} as a supplementary measure to \textbf{Diagonality Index} which is discussed in \textbf{Section 3.4}. Dispersion Index aims to measure to measure the segregation tendency based on the Normalised Interaction Matrix as Recalling that \(N_{i,j}\) is a 10 by 10 matrix with non-negative entries \(a_{ij} \geq\ 0\),  the summation over all entries can be written as \(S =\ \sum_{i=1}^{i}\sum_{j=1}^{j}a_{ij}\). Let \(b=({1,\ 2,\ ...\ ,10})\) be a threshold and for each threshold \(b\) we sum up the diagonal and the off diagonals up to \(S_b\ =\ \sum_{i_b=1}^{i}\sum_{j_b=1}^{j}a_{i_bj_b}\) where \(i_{b},j_{b}\) belong to the first \(b\) diagonals (in both directions). It returns \(S_1\) as the trace, and \(S_c= S\) as the sum of all elements. The absolute dispersion measure is proposed as \\

\begin{align}
\ D_b=\ \frac{S_b}{S}\
\end{align}

and defined as the function of \(b\). Consequently, the presence of large values concentrated around the diagonal contributes to sharp increase along \(b\). In contrast, we expect marginal increment as we increase \(b\) in the case of highly homogeneous matrix  \(N_{i,j}\). \\

To capture the extent that such values differs from by chance, we compute a reference \(D_{rb}\). In this regard, a reference homogeneous matrix \(Mr_{i,j}\) is constructed where all entries are constant. The absolute dispersion measure of the reference matrix \(Dr_b\) is computed likewise. The dispersion in the matrix \(N_{i,j}\) relative to a homogeneous reference point \(Mr_{i,j}\) is measured by \\

\begin{align}
\Delta D = Db - Drb\
\end{align}

retaining the area between the curves of \(D_b\) and \(D_{rb}\) as a function of \(b\). \\

The index value ranges from 0 to 1. The lower boundary indicates complete heterogeneous mixing in which users visit places across socioeconomic status and the upper boundary pinpoints complete homogeneous mixing in which users’ mobilities are concentrated within their own socioeconomic status. Dispersion index provides consistent result with  diagonality index as revealed in Fig. 8.\\

\begin{figure*}[!htb]
 \centering
    \includegraphics[width=0.8\linewidth]{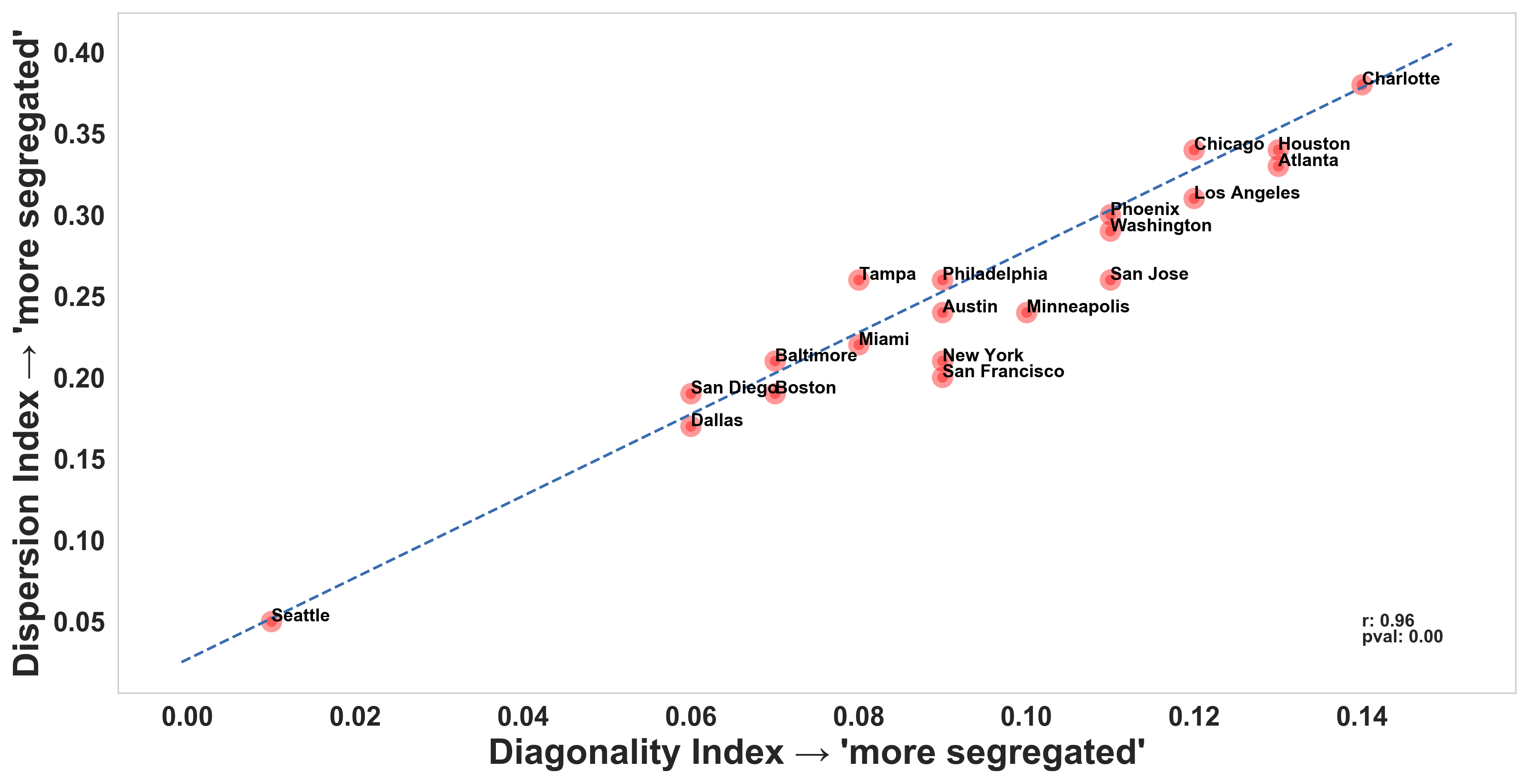}
	\caption{Homophily mixing tendency presented as 2 dimensional measure: Diagonality and Dispersion Index. Charlotte and Seattle seem to be outliers because the two are located away from the rest. Houston has the highest main-homophily mixing tendency among other cities, as well as the strongest neighbouring-homophily mixing. New York is less segregated, while Dallas is even far from being homophily mixing.}
\label{Fig:G1}       
\end{figure*}

\end{document}